\documentclass[ds]{iosart2x}

\makeatletter
\let\numberlines@hook\relax
\makeatother

\usepackage{framed}
\usepackage[normalem]{ulem}
\usepackage{lipsum}                     
\usepackage{xargs}                      
\usepackage[pdftex,dvipsnames]{xcolor}  
\usepackage[colorinlistoftodos,prependcaption,textsize=tiny]{todonotes}
\usepackage[ruled,longend]{algorithm2e}

\newenvironment{widequotation}{\list{}{\listparindent 1.5em \itemindent\listparindent
		\rightmargin 0pt \parsep 0pt plus 1pt}\item\relax}
{\endlist}
\def\signed#1{{\leavevmode\unskip\nobreak\hfil\penalty50\hskip2em
		\hbox{}\nobreak\hfil\raise-1pt\hbox{#1}%
		\parfillskip=0pt \finalhyphendemerits=0 \endgraf}}
\newsavebox\mybox

\usepackage{xcolor}
\usepackage{amssymb}
\usepackage{hyperref}

\usepackage{natbib}

\pubyear{0000}
\volume{0}
\firstpage{1}
\lastpage{1}

\begin{document}

\begin{frontmatter}

\title{Reducing the Effort for Systematic Reviews in Software Engineering}
\runningtitle{Reducing the Effort for Systematic Reviews in Software Engineering}


\author[A]{\inits{F.}\fnms{Francesco} \snm{Osborne}\ead[label=e1]{francesco.osborne@open.ac.uk}\ead[label=o1,orcid]{0000-0001-6557-3131}%
\thanks{Corresponding author. \printead{e1}.}},
\author[B]{\inits{H.}\fnms{Henry} \snm{Muccini}\ead[label=e2]{henry.muccini@univaq.it}\ead[label=o2,orcid]{0000-0001-6365-6515}},
\author[C]{\inits{P.}\fnms{Patricia} \snm{Lago}\ead[label=e3]{p.lago@vu.nl}\ead[label=o3,orcid]{0000-0002-2234-0845}},
and
\author[D]{\inits{E.}\fnms{Enrico} \snm{Motta}\ead[label=e4]{enrico.motta@open.ac.uk}\ead[label=o4,orcid]{0000-0003-0015-1952}}
\runningauthor{F. Osborne et al.}

\address[A]{Knowledge Media Institute, \orgname{The Open University},
\cny{UK}\printead[presep={\\}]{e1,o1}}
\address[B]{DISIM Department, \orgname{University of L'Aquila},
\cny{Italy}\printead[presep={\\}]{e2,o2}}
\address[C]{Department of Computer Science, \orgname{Vrije Universiteit Amsterdam},
\cny{The Netherlands}\printead[presep={\\}]{e3,o3}}
\address[D]{Knowledge Media Institute, \orgname{The Open University},
\cny{UK}\printead[presep={\\}]{e4,o4}}


\begin{abstract}
\textit{Context.} 
Systematic Reviews (SRs) are means for collecting and synthesizing evidence from the identification and analysis of relevant studies from multiple sources. 
To this aim, they use a well-defined methodology meant to mitigate the risks of biases and ensure repeatability for later updates. SRs, however, involve significant effort.
\\
\textit{Goal.} The goal of this paper is to introduce a novel methodology that reduces the amount of manual tedious tasks involved in SRs while taking advantage of the value provided by human expertise.\\
\textit{Method.} Starting from current methodologies for SRs, we replaced the steps of keywording and data extraction with an automatic methodology for generating a domain ontology and classifying the primary studies. This methodology has been applied in the Software Engineering sub-area of Software Architecture and evaluated by human annotators.
\\
\textit{Results.} The result is a novel Expert-Driven Automatic Methodology, EDAM, for  assisting researchers in performing SRs. EDAM combines ontology-learn\-ing techniques and semantic technologies with the human-in-the-loop. The first (thanks to automation) fosters scalability, objectivity, reproducibility and granularity of the studies; the second allows tailoring to the specific focus of the study at hand and knowledge reuse from domain experts. We evaluated EDAM on the field of Software Architecture against six senior researchers. As a result, we found that the performance of the senior researchers in classifying papers was not statistically significantly different from EDAM.
\\
\textit{Conclusions.} Thanks to automation of the less-creative steps in SRs, our methodology allows researchers to skip the tedious tasks of keywording and manually classifying primary studies, thus freeing effort for the analysis and the discussion. 
\end{abstract}


\end{frontmatter}

\section{Introduction} \label{sec:intro}

Understanding the state-of-the-art in research provides the foundation for building novelty. In particular, in Software Engineering topic areas, the acquisition of knowledge for this understanding follows a clear path: started with informal reviews and surveys, it is moving towards systematic searches of the literature. \citet{kitchenhamTR2004} clearly explains the reasons, the importance, and the advantages and disadvantages of using systematic reviews instead of informal ones. Various studies (e.g., \citep{Da_Silva2014,tertiaryAli}) reveal the growing interest in systematic literature reviews and systematic mapping studies \citep{Wohlin2013}. A number of articles and books have been  written on how to perform such systematic studies \citep{kitchenham2007guidelines,wohlin2013systematic,Wieringa2014book}.

A Systematic Review (SR) is {\em ``a means of evaluating and interpreting all available research relevant to a particular research or topic area or phenomenon of interest''} \citep{kitchenhamTR2004}. 
Given a set of research questions, and by following a systematically defined and reproducible process, a SR helps selecting primary studies that contribute to provide an answer to them. Used in combination with keywording~\citep{mapping_se}, a SR supports the systematic elicitation of an ontological classification framework~\citep{petersen2015guidelines}.
In this paper we focus specifically on the field of Software Engineering, but systematic literature reviews and mapping studies are used in several research fields, such as Biomedics~\citep{chinapaw2011relationship}, Robotics~\citep{benitti2012exploring}, Artificial Intelligence~\citep{raza2015review}, Human-Computer Interaction~\citep{mannocci2019evolution}, Psychology~\citep{richards2012computer}, an many others.

A SR can help researchers and practitioners in creating a complete, comprehensive and valid picture of the state-of-the-art about a given theme when the search-space is bounded (e.g., when the search query returns few thousands of articles to scrutinize). However, it falls short when used to investigate the state-of-the-art on an entire research area (e.g., Software Architecture) where the returned entries are  hundreds of thousands  - hence clearly unmanageable. As reported by \citet{Vale2016128} while investigating the state-of-the-art of the Component-based Software Engineering area through an SR, a {\em ``\dots manual search} [restricted only to the most relevant journals and conferences related to the CBSE area] {\em was considered as the primary source, given the infeasibility of analyzing all studies collected from automatic search''}. Still, they had to select, read, and thoroughly analyze 1,231 primary studies. 

In contrast to manually run SRs, several state of the art automated methods allow classifying a document in a certain category or topic~\citep{blei2003latent,mendes2011dbpedia,alghamdi2015survey,schultz1999topic}.
Unfortunately, most current techniques suffer from limitations that make them unsuitable for  systematic reviews. The approaches which exploit keywords as proxy for research areas are unsatisfactory, as they fail to distinguish research topics from other terms that can be used to annotate papers (e.g., ``user case'', ``scalability'') and  to  take advantage of the relationships that hold between research areas (e.g., the fact that ``Software Architecture'' is a sub-area of ``Software Engineering''). Probabilistic topic models (e.g., Latent Dirichlet Allocation~\citep{blei2003latent}) are also unsuitable for this task  since they produce cluster of terms that are not easy to map to  research areas~\citep{osborne2013exploring}. Crucially, it is often unfeasible to integrate these topic detection techniques with the needs and the knowledge of human experts. Another alternative is to apply entity linking techniques~\citep{mendes2011dbpedia} to map papers to relevant entities in knowledge base. Unfortunately, we currently lack good granular and machine readable representation of research areas in many domains which could be used to this end. 

Current techniques have complementary limitations when investigating the state-of-the-art of an entire research area: on the one hand side, SRs are {\em ``human-intensive''}, as they require domain experts to invest a large amount of time to carry out manual tasks; on the other side, automated techniques keep the {\em humans ``out of the loop''}, while human expertise is critical for the more conceptual analysis tasks.

This paper proposes an {\em expert-driven automatic methodology} (EDAM) for assisting systematic reviews that, while recognizing the essential value of human expertise, limits the amount of tedious tasks the expert has to carry out. Our methodology contributes with 1) automatically extracting an ontology of relevant topics, related to a given research area; 2) using experts to refine this knowledge base;
3) exploiting this knowledge base for classifying relevant papers that may be then further validated/analyzed by experts, and for computing research analytics. 
We demonstrate EDAM in the field of Software Architecture, but it can be easily applied to other research areas as well.

Naturally, the ability of domain experts to analyse the research dynamics emerging from primary studies and to distill the most important lessons and trends is still crucial. Therefore, our aim is not to fully automatize the process, but to assist domain experts by automatically generating data-driven analytics in order to free time and resources for the analysis phase. 
 
In summary, our contributions are: 

\begin{itemize}
\item a novel methodology for supporting ontology-driven systematic reviews, which involves both automatic techniques and human experts;
\item an implementation of this methodology which exploits the Klink-2 algorithm for generating the domain ontology in the field of Software Architecture;
\item an illustrative analysis of the Software Architecture trends;
\item an evaluation involving six human annotators, which shows that the classification of primary studies yielded by the proposed methodology is comparable to the one produced by domain experts (p=0.77).
\item an automatically generated ontology of Software Engineering,  which could support further systematic reviews in the field\footnote{\url{http://rexplore.kmi.open.ac.uk/data/edam/SE-ontology.owl}}.

\end{itemize}

The rest of the paper is structured as follows. Section \ref{sec:rw} introduces related works on systematic studies. 
Section \ref{sec:why} provides an overview of some preliminary evidence of the benefits brought by using EDAM to assist a mapping study.
Section \ref{sec:method} then presents the EDAM methodology and its application to the research area of Software Architecture.
This experiment is discussed and evaluated in Section \ref{sec:discussion}, which also present a comparison of several approaches for classifying research papers. Finally, in Section~\ref{sec:conclusion} we discuss the main implications of our study and outline future directions of research.

\section{Related Work}  \label{sec:rw}

There are many guidelines for, and reports on, carrying out systematic studies in Software Engineering. Among them, we could identify a few aimed at supporting or improving the underlying process. In our perspective, they all enable researchers to focus more on the most creative steps of a systematic study by removing what is referred to as {\em manual work}.

With a motivation similar to ours, i.e. to improve the search step in systematic studies in Software Engineering research,
\citet{Octaviano2015-xr} propose a strategy that automates part of the primary study selection activity.
\citet{Mourao2017} present a preliminary assessment of a hybrid search strategy for systematic literature reviews that combines database search and snowballing to reduce the effort due to searches in multiple digital libraries.
\citet{Kuhrmann2017} provide recommendations specifically for the general study design, data collection, and study selection procedures.
\citet{Zhang2011}, in turn, systematically select and analyze a large number of SRs. Their results have been then used to define a quasi-gold standard for future studies. In their validation, they were able to improve the rigor of the search process and provide guidelines complementing the ones already in use. 

\citet{runeson17} propose a machine learning approach that classifies papers for SRs by leveraging human experts, who iteratively validate set of publications produced by a classifier. Conversely, EDAM does not require experts to manually examine research papers, but only to review a taxonomy of research areas.

The need for guidelines in conducting empirical research has been addressed in other types of empirical studies, too.
\citet{DeMello2016} focus on opinion surveys and provide guidelines (in the form of a reference framework) aimed at improving the representativeness of samples. Also on opinion surveys,
\citet{Molleri2016} provide recommendations based on an annotated bibliography instead. 

Another interesting work by 
\citet{Felizardo2016} investigates how the use of forward snowballing can considerably reduce the effort in updating SRs in Software Engineering. Based on this result, complementing our method with automated forward snowballing suggests a very promising direction for future works as it could further reduce the effort for identifying relevant primary studies.

\citet{Marshall2015} carried out an interview survey with experts in other domains (i.e. healthcare and social sciences) with the aim to identify tools that are generally used, or desirable, to ease which steps in systematic studies, and transfer the best practices to the Software Engineering domain. Among the results, data extraction and automated analysis emerge as top requirements for reducing the workload. In a similar vein, \citet{IST-SLRtoolsupport} followed by \citet{Al-Zubidy2017-np} consulted Software Engineering researchers conducting SRs to identify and prioritize the necessary SR tool features. The results identified {\em search \& study selection} as the most desirable feature. Our work addresses the needs identified by both \citet{Marshall2015} and \citet{IST-SLRtoolsupport}.

The idea of using ontologies for supporting SRs was discussed by few papers, but did not receive much attention.
\citet{de2007scientific} introduced the Scientific Research Ontology, a resource to organize the knowledge generated from SR. This ontology offers a conceptual framework with the aim of fostering the consistency between different studies, but does not directly assist the tasks involved in SR, such as the extraction of primary studies.
\citet{sun2012towards} discussed the use of ontologies for supporting key activities in SRs and presented an experiment in which they automatically classified primary studies by means of COSONT, an ontology of methods for cost estimation. Unfortunately, their approach still required to manually check hundreds of papers and the COSONT ontology was quite simplistic, being an handcrafted list of methods with no hierarchical structure. This is a common issue with manually generated ontology of research concepts, which are usually costly to produce, coarse-grained, and slow to evolve~\citep{osborne2015klink}.
Conversely, EDAM takes advantage of recent ontology learning techniques to automatically generate complex multi-level ontologies (e.g., the SE ontology presented in this paper includes 956 topics and 5,461 relationships), exploits the resulting taxonomic structure to classify the primary studies, and does not require experts to  manually review a large number of papers.

\section{An Overview of the Benefits of Automatic SRs} \label{sec:why}

Before entering the details of the EDAM methodology, this section provides an overview of the benefits such an automatic SR methodology can bring with respect to more traditional, manual SRs carried out according to predefined protocols.

We all agree that manual SRs based on well-defined systematic protocols help reducing (but not fully removing) subjective biases in the selection of the studies. They however are by and large unfeasible in reviewing a too large dataset (i.e. when the number of scientific publications is too large to be manually processed by the researcher).

In a similar vein, automatic SRs help reducing subjective biases (in this case by {\em implementing} the selection of the studies according to the predefined systematic protocol). Differently, they pose no limitation in terms of the size of the dataset of publications.

In our earlier work~\citep{Wolfram2017-ts} we challenged these limitations and benefits by applying the automatic study selection to a manual SR carried out beforehand by other researchers~\citep{TR-SLR-sustainab}. In this way, we could compare and contrast the results of the manual SR with the results of our automatic SR.
In this earlier work, we have studied the field of software sustainability within the Software Engineering domain. While at the time the EDAM methodology was not yet fully developed, we did use the same ontology-learning algorithms and a preliminary version of the ontology for the Software Engineering domain.

\begin{figure}[htpb]
\centering
\includegraphics[scale=0.5]{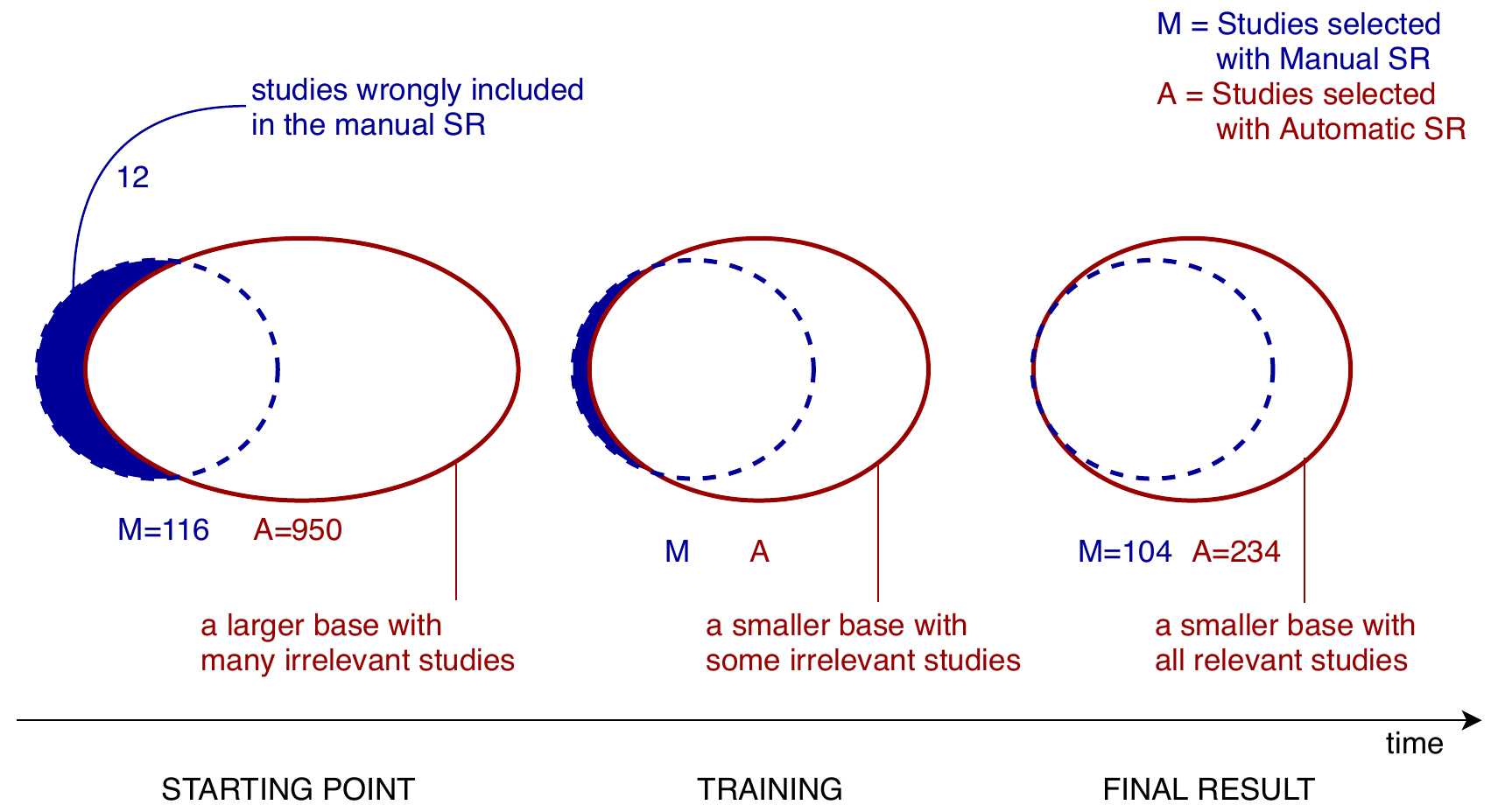}
\caption{Some evidence on the Benefits of Automated SRs}
\label{fig:whyedam}
\end{figure}

The observations gathered during this experiment are illustrated in Fig.~\ref{fig:whyedam}, where we represented the primary studies selected manually (see the left-hand circles) and those selected automatically (see the right-hand ovals). The experiment underwent three phases:

\begin{description}
\item[{\bf Starting point:}] The already-completed manual SR had selected 116 primary studies. Before training the algorithm and tuning the domain ontology, from the Scopus dump of scientific publications we automatically selected 950 studies. While our automatic methodology is able to handle seamlessly any size of the base of publications, the selected studies did initially include a very large number of false positives. However, it did also uncover that 12 studies selected in the manual SR where wrongly included. {\em {\bf Observation \#1:} in spite of systematic selection criteria and the involvement of multiple researchers, human errors in the manual study selection is still possible.}

\item[{\bf Training:}] By treating the 104 primary studies (from the manual SR) as pilot studies, we trained our domain ontology and learning algorithm to automatically select the primary studies. {\em {\bf Observation \#2:} Automatic SR is able to automatize the selection criteria of systematic reviews while handling any size of the initial dataset of scientific publications.} As discussed in Section~\ref{sec:eval1}, the domain ontology is able to classify the primary studies as correctly as the human experts do, without needing further training. As such, the domain ontology can be reused for any study in the domain of Software Engineering.

\item[{\bf Final result:}] The final result of the automatic selection converged to 234 studies which included the 104 pilot studies and {\em correctly} identified additional 130 studies that were missing in the original manual SR. {\em {\bf Observation \#3:} By handling a much larger base of publications, automatic SRs are able to uncover primary studies that are missed by manual SRs where such scale is unfeasible.}
\end{description}

\section{An Expert-Driven Automatic Methodology}  \label{sec:method}

We propose a novel expert-driven automatic methodology (EDAM) for assisting systematic reviews like systematic literature reviews and mapping studies. EDAM allows to automatize the steps that are the most time and effort consuming while requiring the least creativity, such as \textit{selection of relevant papers}, \textit{keywording}, and \textit{creation of a classification schema}~\citep{petersen2015guidelines}, by exploiting ontology learning techniques and semantic technologies to foster scalability, objectivity, reproducibility, and granularity of the study (further discussed in Section~\ref{sec:implications}). It also supports the generation of research trends, which are typical of data synthesis in mapping studies. In this paper, we illustrate how EDAM can support mapping studies, even though it can be evidently exploited in systematic literature reviews, too.

\begin{figure}[htpb]
\centering
\includegraphics[scale=0.3]{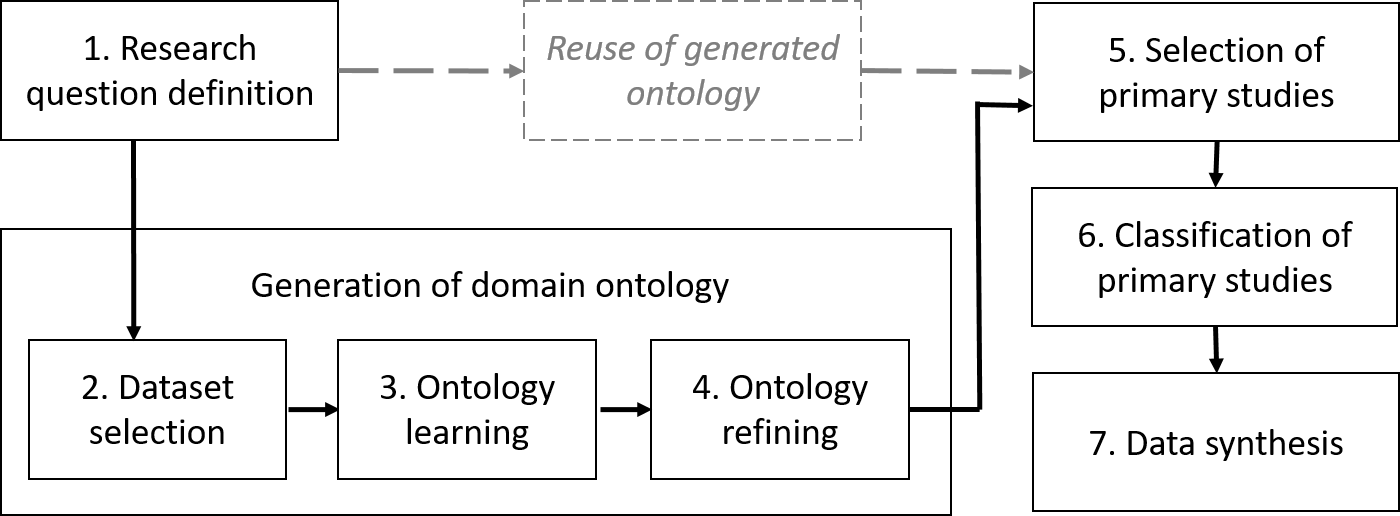}
\caption{Steps of a systematic mappings adopting the EDAM methodology. The gray-shaded elements refer to the alternative step of reusing the previously generated ontology.  }
\label{fig:edam}
\end{figure}	

\begin{figure}[htpb]
\centering
\includegraphics[scale=0.55]{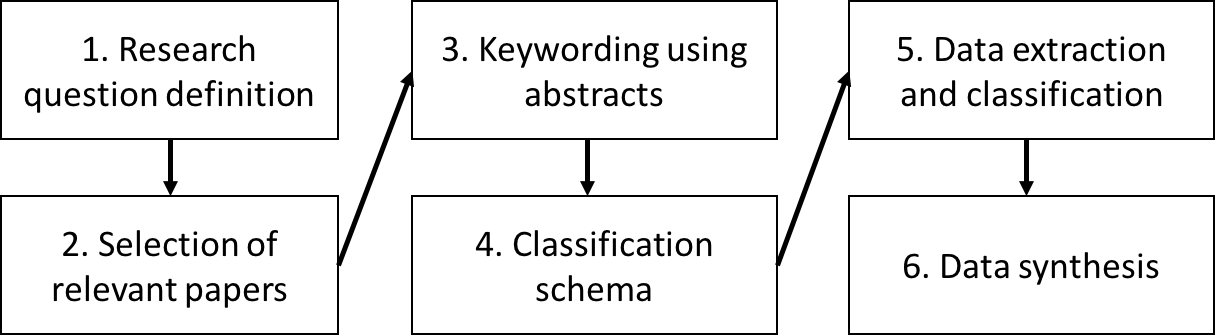}
\caption{Classic steps of systematic mappings (inspired by~\citep{petersen2015guidelines})}
\label{fig:ms}
\end{figure}

Figure \ref{fig:edam} shows the steps of a mapping study using EDAM in contrast with the steps of a classic (manual) methodology - shown in Figure \ref{fig:ms}. The main difference is that in the classic methodology the researchers first select and analyze each primary study (steps 2-3) and then produce a taxonomy to classify them (step 4). When 
assisted by EDAM, instead, the researchers first use ontology learning methods over large scholarly datasets to generate an ontology of the field (steps 2-3), then refine the ontology with the help of domain experts (step 4), and finally exploit this knowledge base to automatically select and classify the primary studies (steps 5-6).  

An alternative solution for steps 2-4 (Generation of domain ontology) is the reuse of an ontology crafted by a previous study with the same scope. Indeed, in the study discussed in Section \ref{sec:methodologyimplementation} we have generated an ontology of Software Engineering (SE) research topics, with the hope that it will be re-used by the research community.

In Section \ref{sec:generalmethodology}, we describe EDAM and discuss its advantages over a classic methodology. In Section \ref{sec:methodologyimplementation}, we exemplify the application of EDAM specifically aimed at identifying publication trends of the Software Architecture research area in the specific SE domain.

\subsection{EDAM Description} \label{sec:generalmethodology}
A SR assisted by EDAM is organized along the following steps (ref. Figure~\ref{fig:edam}). \\

\textbf{1. Research question definition.} The researchers performing the study state the research questions (RQs). These will affect the aim of the study and thus its steps. It should be noted that EDAM is applicable only to research questions that could be answered by classifying publications, authors, venues, and other entities according to the ontology for producing relevant analytics. Other research questions should be addressed according to the standard methodology~\citep{petersen2015guidelines}. \\

\textbf{2. Dataset selection.} The researchers select a dataset on which to apply the chosen ontology learning technique (further elaborated in step 3) for generating the domain ontology that will be used to select and classify the primary studies. The most important characteristic of this dataset is that it must be unbiased with respect to the focus of the study. For example, if the study wants to uncover the trends in research areas (e.g., Software Architecture), the dataset should not be biased with respect to any area in the domain (e.g., Software Engineering in our case). A good strategy to select unbiased datasets is considering either a full scholarly dataset of a very high-level field (e.g., all the Computer Science papers in Microsoft Academic Search\footnote{\url{http://academic.research.microsoft.com}} or in Scopus\footnote{\url{https://www.scopus.com/}}) or a dataset including all the papers published in the main conferences and journals of the domain under analysis. In recent years, universities, organizations, and publishing companies have released an increasing number of open datasets that could assist in this task, such as CrossRef\footnote{\url{https://www.crossref.org/}}, SciGraph\footnote{\url{https://scigraph.springernature.com/explorer/downloads/}}, OpenCitations\footnote{\url{http://opencitations.net}}, DBLP\footnote{\url{http://dblp.uni-trier.de}}, Semantic Scholar\footnote{\url{https://www.semanticscholar.org/}}, and others. \\

\textbf{3. Ontology learning.} The dataset is processed by an ontology learning technique that automatically infers an ontology of the relevant concepts. 

We strongly advocate the use of an ontology learning technique that generates a full domain ontology and represents it with Semantic Web standards, such as the Web Ontology Language (OWL)\footnote{\url{https://www.w3.org/OWL/}}). 
The main advantage of adopting an ontology in this context is that it allows for a more comprehensive representation of the domain since it includes, in addition to hierarchical relationships, also other kinds of relationships (e.g., \textit{sameAs}, \textit{partOf}), which may be critical for classifying the primary studies. For example, an ontology allows to explicitly associate to each category a list of alternative labels or related terms that will be used in the classification phase. In addition, ontology learning techniques can infer very structured multi-level ontologies~\citep{osborne2015klink}, and thus describe the domain at different levels of granularity. 

The task of ontology and taxonomy learning was comprehensively explored over the last 20 years. Therefore, the researcher can choose among a variety of different approaches for this step, including:
\begin{itemize}
  \item statistical methods for deriving taxonomies from keywords~\citep{sanderson1999deriving,liu2012automatic};
  \item natural language processing approaches, e.g., FRED~\citep{gangemi2017semantic}, LODifier~\citep{augenstein2012lodifier}, Text2Onto~\citep{cimiano2005text2onto};
 \item approaches based on deep learning, e.g., recurrent neural networks~\citep{petrucci2016ontology};
 \item hybrid ontology learning frameworks~\citep{wohlgenannt2012dynamic};
  \item specific approaches for generating research topic ontologies, e.g., Klink-2~\citep{osborne2015klink}.
\end{itemize}

However, as discussed in the following step, researchers may also choose to skip this step and re-use a compatible ontology from a previous study.

It is useful to clarify why we suggest the adoption of an ontology learning approach, rather than the adoption of one of the currently available research taxonomies, such as the ACM computing classification system\footnote{\url{http://www.acm.org/publications/class-2012}}, the Springer Nature classification\footnote{\url{http://www.nature.com/subjects}}, Scopus subject areas\footnote{\url{https://www.elsevier.com/solutions/scopus/content}}, and the Microsoft Academic Search classification. Unfortunately, these taxonomies suffer from some common issues, which make them unfeasible to support most kinds of SRs. First, they are very coarse-grained and represent wide categories of approaches, rather than the fine-grained topics addressed by researchers ~\citep{osborne2012mining}. Secondly, they are usually obsolete since they are seldom updated. For example, the 2012 version of the ACM classification was finalized fourteen years after the previous version. This is a critical point, since some interesting trends could be associated with recently emerged topics. In third instance, most ontology learning algorithms are not limited to learning research areas, but can be tailored to yield the outputs which are more apt to support a specific analysis.\\

\textbf{4. Ontology refining.} The ontology resulting from the previous step is corrected and refined by domain experts. During this phase, the experts are allowed to 1) delete an existent category, 2) add a new category, 3) delete an existent relationship, 4) add a new relationship. We suggest using at least three domain experts for addressing possible disagreements.

This step is critical for two reasons. First, it may correct some errors in the automatically-generated taxonomy. Secondly, it verifies that the data-driven representation aligns with the domain experts mental model and thus the outcomes will be understandable and reusable by their research community.

Refining a very large ontology is not a trivial task, therefore if the domain comprehends a large number of topics we suggest splitting it in manageable sub branches to be addressed by different experts. Our experience suggests that a taxonomy of about 50 research areas can be reviewed in about 15-30 minutes by an expert of the field. For example, in ~\citep{osborne2015klink} three experts reviewed a Semantic Web ontology of 58 topics in about 20 minutes. In the test study for this paper, three experts took about 20 minutes to examine and produce feedback on a taxonomy of  46 topics (and 71 terms considering synonymous such as ``product line'', ``product-lines'', ``product-line'', which  were clustered automatically by the ontology learning algorithm). In both cases, we represented the ontology as tree diagram in a excel sheet\footnote{See an example at \url{http://tinyurl.com/yal6h3wu}} and included also a list of the most popular terms in the dataset, for supporting experts in remembering all the relevant research topics.

An alternative solution is to provide experts with ontology editors that could be used to directly modify the ontology, such as Protege\footnote{\url{http://protege.stanford.edu}},  NeOn Toolkit\footnote{\url{http://neon-toolkit.org/}},  TopBraid Composer\footnote{\url{http://www.topquadrant.com/products/TB_Composer.html}},  Semantic Turkey\footnote{\url{http://semanticturkey.uniroma2.it/}}, or Fluent Editor\footnote{\url{http://www.cognitum.eu/Semantics/FluentEditor/}}. However, these tools are not always easy to learn and the adoption of a simple spreadsheet may be advisable in most cases. 
Indeed, the annotators who participated in the mapping study of Software Architecture described in the next section, reported that they were able to easily correct and suggest changes in the ontology using this simple solution. In particular, this task was natural to them since the same kind of spreadsheet is typically used in the analysis phase of systematic reviews (e.g., for the keywording step). 
We refer the reader to \citet{sabou2018verifying} for a comprehensive analysis of the verification of domain knowledge by human experts in the field of Software Engineering.

As highlighted by Figure~\ref{fig:edam}, the aim of steps 2-4 is to generate an ontology apt to select and classify relevant papers and ultimately answer the RQs. It follows that these steps could be replaced by the adoption of an ontology previously generated and validated by a previous study with a consistent scope. For example, the ontology about Software Engineering generated for this paper's example study (see Section~\ref{sec:methodologyimplementation}) can be re-used to perform many kinds of mapping studies involving other research areas in SE. Naturally, the ontology may have to be further updated to include the most recent concepts and terms. This solution allows users with no access to vast scholarly databases or no expertise in ontology learning techniques to easily implement an EDAM study.\\

\textbf{5. Selection of primary studies.}
The authors select a dataset of papers and define the inclusion criteria of the primary studies according to the domain ontology and other metadata of the papers (e.g., year, venue, language).
The inclusion criteria are typically expressed as a search string, which uses simple logic constructs, such as  AND, OR, and NOT \citep{aromataris2014constructing}. The search string is then  used to produce the query that will be run over the dataset for selecting the primary studies. 
Some examples of queries include 1) ``all the papers in the dataset published in a list of relevant conferences'' or ``all the papers in the dataset that contain a list of relevant terms from the ontology''. 

In most cases this dataset will be the same or a subset of the one used for learning the domain ontology. However, the authors may want to zoom on a particular set of articles, such as the ones published in the main venues of a field, in a geographical area, or by a certain demography. It is also possible to select a different dataset altogether, since the ontology would use generic topic labels and thus be agnostic with respect to the dataset. A possible reason to do so is the availability of the full text of the studies. Many ontology learning algorithms can be run on massive metadata dataset  (e.g., Scopus, Microsoft Academic Search), but some research questions may require the full text. In this case, the author may want to perform the ontology learning step on the metadata dataset, which is usually larger in size and scope, and then either select a subset composed by publications which are available online or adopt for this phase a second dataset that includes the full text of the articles, such as Core \citep{knoth2012core}. The growth of the Open Access movement \citep{wilkinson2016fair}, which aims at providing free access to academic work, may alleviate this limitation in the following years.\\

\textbf{6. Classification of primary studies.}
The authors define a function for mapping categories to papers based on the refined ontology. This step is important to foster reproducibility since the inclusion criteria (defined in the step 5), the mapping function, and the domain ontology should contain all the information needed for replicating the classification process. The function can also be associated to an algorithmic method (e.g., a machine learning classifier), provided the method is made available and is reproducible. 

The simplest way for mapping categories to papers is to associate to each category each paper that contains the label of the category or of any of its sub-categories. This simple technique for semantically characterizing documents has the advantage of being unsupervised and was applied with good results in a variety of fields, such as topic forecasting \citep{salatino2017topics}, automatic classification of proceeding books \citep{osborne2016automatic},  sentiment analysis \citep{saif2012semantic}, recommender systems \citep{di2012linked}, and many others. Alternative unsupervised methods, which we evaluate in Section \ref{sec:eval2}, include approaches based on TF-IDF~\citep{ramos2003using}, LDA~\citep{blei2003latent}, and word embeddings~\citep{salatino2018csoc2}. 
 
In addition, the authors can choose to create a more complex mapping function which exploits other semantic relationships in the ontology (e.g., \textit{relatedTerm}, \textit{partOf}). \\ 

\textbf{7. Data synthesis.} According to the RQs, this step may be automatic, semi-automatic or manual. Some straightforward analytics (e.g., the number of publications or citations over time) can be computed completely automatically by counting the previously classified papers or summing their number of citations. Other more complex analyses may require the use of machine learning techniques or the (manual) intervention of human experts. Starting from the groundwork formed by our research, a full analysis of the possible kinds of data synthesis and the way to automatize them will constitute interesting future works beneficial for the whole research community.\\

Overall, motivated by the need to reduce the amount of manual tedious tasks involved in SRs, 
{\bf EDAM offers four main advantages over a classic methodology}. {\bf First,} human experts are not required to manually analyze and classify primary studies, but they simply have to refine the ontology, choose the inclusion criteria, and define a mapping function for associating papers to categories in the ontology. This allows researchers to carry out large scale studies that involve thousands of research papers with relative ease. {\bf Secondly,} since the domain ontology is created with a data-driven method, it should reflect the real trends of the primary studies, rather than arbitrary human decisions about which keywords to annotate and aggregate, even if the refinement step may still introduce a degree of arbitrariness. {\bf Third,} the use of a formal machine-readable ontology language for representing the domain taxonomy should foster the reproducibility of the study and allow authors with no expertise in data science to perform studies using previously generated ontologies. {\bf Fourth,} this methodology allows researchers to produce and exploit complex multi-level ontologies, rather than the simple two-level classifications used by many studies~\citep{Vale2016128}.

Naturally, EDAM is suitable for research questions that can be automatized by the ontology-driven classification process previously described, or that aim at giving an overview of the state-of-the-art or state-of-practice on a topic~\citep{wohlin2012experimentation} by analysing all of the relevant research contributions in a specific research area.
We will discuss further this and other limitations in Section \ref{sec:limitations}.

\subsection{EDAM Application}
 \label{sec:methodologyimplementation}

With the aim of presenting a reproducible pipeline and showing how EDAM can be applied, we present here an example as part of a possible systematic mapping study assisted by EDAM in the Software Architecture research area. We chose to study the research trends in this area, since trend analysis is typical of mapping studies~\citep{wohlin2012experimentation} and it is one of the tasks that can be automatized by EDAM.

In the following, we describe how we instantiated the study example assisted by EDAM and discuss the specific technologies used to implement it. The data necessary for reproducing this study and using this same pipeline on other fields are available at 
\url{https://doi.org/10.5281/zenodo.2653924}.\\

\textbf{1. Research question definition.} We wanted to focus on a task that is often addressed by mapping studies and could be completely automatized. Therefore our RQ is: ``What are the trends of the main research topics of Software Architecture?''.\\

\textbf{2. Dataset selection.} We selected all papers in a dump of the Scopus dataset about Computer Science in the period 2005-2013. The Scopus dataset we were given access by Elsevier BV includes papers in 1900-2013 interval, but the number of relevant articles before 2005 was too low to allow a proper trend analysis. Each paper in this dataset is described by title, abstract, keywords, venue, and author list.\\

\textbf{3. Ontology learning.} We applied the Klink-2 algorithm~\citep{osborne2015klink} on the Scopus dump   for learning an ontology representing the main 'Software Architecture' research area in SE.

Klink-2 is an algorithm that generates an ontology of research topics by processing scholarly metadata (titles, abstracts, keywords, authors, venues) and external sources (e.g., DBpedia, calls for papers, web pages). 
In particular, Klink-2 periodically produces the Computer Science Ontology (CSO)\footnote{\url{http://cso.kmi.open.ac.uk/}}~\citep{salatino2018computer} that is currently used by Springer Nature for classifying proceedings in the field of Computer Science~\citep{osborne2016automatic}, such as the well-known Lecture Notes in Computer Science series\footnote{\url{http://www.springer.com/gp/computer-science/lncs}}. The ontologies produced by Klink-2 use the Klink data model\footnote{\url{http://technologies.kmi.open.ac.uk/rexplore/ontologies/BiboExtension.owl}}, which is an extension of the BIBO ontology\footnote{\url{http://purl.org/ontology/bibo/}} that in turn builds upon SKOS\footnote{\url{https://www.w3.org/2004/02/skos/}}. This model includes three semantic relations: \textit{relatedEquivalent}, which indicates that two topics can be treated as equivalent for the purpose of exploring research data; \textit{skos:broaderGeneric}, which indicates that a topic is a subarea of another one; and \textit{contributesTo}, which indicates that the research outputs of one topic significantly contribute to the research into another. In the following, we make use of the first two relationships for classifying studies according to their research topics.

\begin{algorithm}[H]
\label{alg:klink-2}
\LinesNumbered
\SetKwInOut{Input}{Input}\SetKwInOut{Output}{Output}

\Input{List of keywords $keywords$, Metadata $metadata$ }
\Output{Ontology $ontology$}
\BlankLine

 relationships=\{\}; 
\BlankLine
\While{some keywords yet to process}{
\ForEach{k1 \textbf{in} keywords}{
candidates = GetCandidates(k1, metadata)\;
\ForEach{k2 \textbf{in} candidates}{
    relationship = InferRelationship(k1, k2, metadata, relationships);
}
}
\BlankLine
relationships = RemoveLoops(relationships)\;
new.keywords = MergeAndSplitKeywords(keywords, metadata, relationships)\;
keywords = AddNewKeywords(new.keywords)\;
}
keywords = FilterTopics(keywords, metadata, relationships)\;
\BlankLine
ontology = GenerateSemanticRelationships(relationships)\;
return(ontology)\;
\caption{The Klink-2 algorithm.}
\end{algorithm}

In Algorithm~\ref{alg:klink-2}, we report the pseudocode of Klink-2. The algorithm takes as input a set of keywords and investigates their relationships with the set of their most co-occurring keywords. Klink-2 infers a sub-topic relationship between keyword $x$ and $y$ by means of two metrics: i) $H_R (x,y)$, which uses a semantic variation of the subsumption method; ii) $T_R (x,y)$, which uses temporal information to do the same.
$H_R (x,y)$ is computed according to the following formula:
$$H_R (x,y)= \left(\frac{I_R (x,y)}{I_R (x,x)} - \frac{I_R (y,x)}{I_R (y,y)}\right) c_R (x,y) ~ n(x,y)$$
where $I_R (x,y)$ is the number of elements associated with both x and y according to relation $R$ (e.g., number of co-occurrences in research papers), $\frac{I_R (x,y)}{I_R (x,x)}$ is the conditional probability that an element associated with keyword $x$ will be associated also with keyword $y$, $n_R (x,y)$ is the Levenshtein distance between the two keywords normalized by the length of the longest one, and $c_R (x,y)$ is the cosine similarity between the two vectors in which each index represents a keyword $k$ and the value is the number of time $x$ or $y$ co-occurred with $k$ in a certain context.

$T_R (x,y)$ is a temporal version of $H_R (x,y)$, which weighs more the information associated with the first years of $x$. It is useful to detect the cases in which the relationship between two terms fade because their association has become implicit (e.g., Artificial Intelligence and Machine Learning). $T_R (x,y)$ is calculated using a variation of formula (1) in which $I_R (x,y)$  is computed by weighting the intensity of the relationships in each year according to the distance from the debut of x. The weight is computed as $w(year, x)= (year - debut(x) +1)^{–\gamma}$, with $\gamma >0$ ($\gamma =2$ in the implementation used for this paper).
A hierarchical relationship is inferred whenever $H_R (x,y)$ or $T_R (x,y)$ are higher then a certain threshold (0.25 in the implementation used for this study).

After inferring the hierarchical relationships, Klink-2 removes loops in the topic network (instruction $\#9$), merges similar keywords and splits ambiguous keywords associated to multiple meanings (e.g., 'Java'). The keywords produced in this step are added to the initial set of keywords to be further analysed in the next iteration and the while-loop is re-executed until there are no more keywords to be processed. Finally, Klink-2 filters the keywords considered 'too generic' or 'not academic' according to a set of heuristics (instruction $\#13$) and generates the triples describing the ontology. 

Klink-2 was evaluated on a gold standard ontology including 88 research topics in the field of Semantic Web, which was manually generated by three senior researchers. It significantly outperformed the alternative algorithms ($p=0.0005$), yielding a precision of 86\% and a recall of 85.5\%. More details about Klink-2 and its evaluation can be found in~\citet{osborne2015klink}. 

We selected Klink-2 among the other previously discussed solutions for a number of reasons. First, it is the only approach to our knowledge that was specifically designed to generate taxonomy of research areas. Secondly, it was already integrated and evaluated on a dump of the Scopus dataset, which we adopted in this study, yielding excellent performance on the fields of artificial intelligence and semantic web ~\citep{osborne2015klink}. In third instance, it permits to define a number of pre-determinate relationships as basis for a new taxonomy. In particular, a human user can define a subsumption relation (i.e.,  \textit{skos:broaderGeneric}), a \textit{relatedEquivalent} one, or specify that two concepts should not be in any relationships. This functionality allows us to easily incorporate expert feedback in the ontology learning process. Therefore, the next iterations of the ontology will benefit from the knowledge of previous reviewers. 
We ran Klink-2 on the selected dataset, giving as initial seed the keyword ``Software Engineering'' and generated an OWL ontology of the field including 956 concepts and 5,461 relationships. We then selected the sub-branch of Software Architecture comprising 46 research areas and 71 terms (some research areas have multiple labels, such as ``component based software'' and ``component-based software''). \\

\textbf{4. Ontology refining.} 
We generated a spreadsheet, containing  the  Software Architecture (SA) ontology as a  tree diagram\footnote{\url{http://tinyurl.com/yal6h3wu}}. In this representation each concept of the ontology was illustrated by its level in the taxonomy, its labels, and the number of papers annotated with the concepts.  We also included a list of the 500 more popular terms in the papers that contained the keywords ``Software Architecture'' and ``Software Engineering'', to assist the experts in remembering other concepts or terms that the algorithm may have missed. 

We sent it to three senior researchers and asked them to correct the ontology as discussed in Section~\ref{sec:generalmethodology}.  The task took about 20 minutes and produced three revised spreadsheets. The feedback from the experts was integrated in the final ontology\footnote{\url{http://rexplore.kmi.open.ac.uk/data/edam/SE-ontology.owl}}. In case of disagreement we went with the majority vote.

The most frequent feedback regarded: 1) the deletion sub-areas that were incorrectly classified under SA (e.g., ``software evolution''), 2) the introduction of sub-areas that were neglected by Klink-2 (e.g, ``architecture concerns''), and 3) the inclusion of alternative labels for some category (e.g., alternative ways to spell ``component-based architecture'').\\

\textbf{5. Selection of primary studies.} 
We then selected from the initial Scopus dump two datasets of primary studies to investigate the SA area: 1) \textbf{DSA} (Dataset SA, 3,467 publications), including all papers in the Scopus dataset that contain the terms ``Software Architectures'' or ``Software Architecture'' and include at least one of the subtopics of Software Architecture in the domain ontology, and 2) \textbf{DSA-MV} (Dataset SA - Main Venues, 1,586 publications), containing all the papers published in a list of well-known conferences and journals in the SE fields and in a particular in the SA area (see Table~\ref{tbl:venuetable}) and including at least one of the sub-topics of SA in the OWL ontology. We considered these two datasets since it may be interesting to analyze the discrepancy between generic SA papers and papers published in the main venues.\\

\textbf{6. Classification of primary studies.} 
We defined the mapping function as follows. A paper was classified under a certain category (e.g., service-oriented architectures) if it contained in the title, abstract or keywords: 1) the label of the category (e.g., ``service-oriented architectures''), 2) a {\em relevantEquivalent} of the category (e.g., ``service oriented architecture''), 3) a {\em skos:broaderGeneric} of the category (e.g., ``microservices''), or 4) a {\em relevantEquivalent} of any {\em skos:broaderGeneric} of the category (e.g., ``microservice''). 
The advantage of this solution is that it allows us to map each category to a list of terms that can be automatically searched in the metadata of the papers. Therefore, the classification step can be handled automatically. In addition, it allows us to associate multiple categories to the same paper.

We chose this straightforward approach instead of other more complex methods based on word embeddings and string similarity~\citep{salatino2018csoc2}, since it is simple to reproduce and yields the best precision, as discussed in Section~\ref{sec:eval2}. There we discuss also some recent approaches~\citep{salatino2018csoc,salatino2018csoc2} yielding a more comprehensive set of topics and therefore a better recall, and illustrate how the choice of the method ultimately depends on the requested tradeoff between precision and recall.

In practice, we indexed titles, abstracts and keywords in an ElasticSearch\footnote{\url{https://www.elastic.co/}} instance and we ran a PHP script that imported the ontology, performed the relevant queries on the metadata, and saved the result in a MariaSQL database\footnote{\url{https://mariadb.org/}}.\\

\begin{figure}[htpb]
\centering
\includegraphics[scale=0.15]{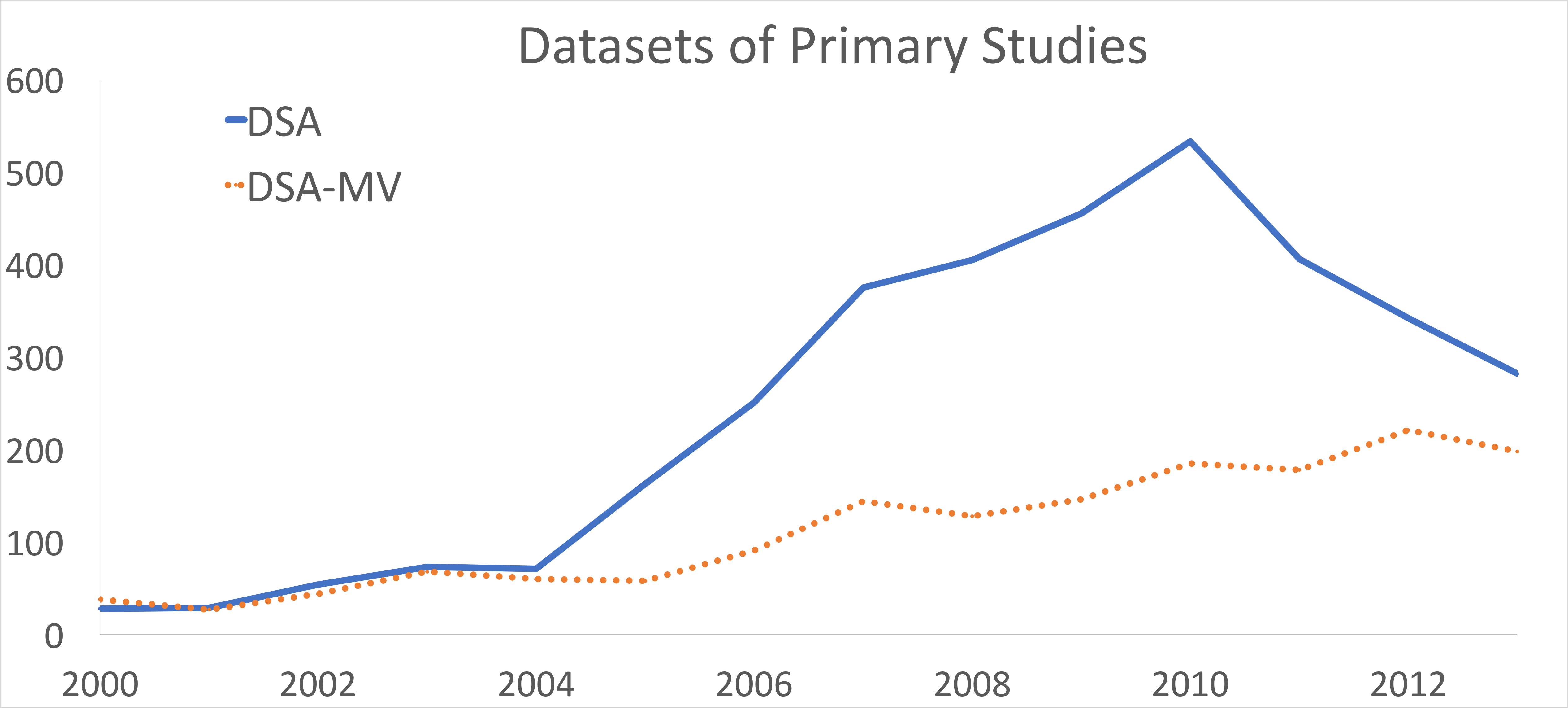}
\caption{Number of publications in DSA and DSA-MV over the years. }
\label{fig:study1}
\end{figure}

\textbf{7. Data synthesis.}
Figure~\ref{fig:study1} shows the number of primary studies in the DSA and DSA-MV datasets. The DSA dataset follows the trend of the ``Software Architecture'' keyword in the Scopus dataset and decreases after 2010. Conversely, the size of DSA-MV grows steadily with the number of relevant conferences and journals. 

We identified the main trends by running a script to count the number of studies about each sub-topic in each year. Since the focus of the paper is the EDAM methodology, rather than a comprehensive analysis on these research sub-areas, we will briefly discuss only the main trends associated with the more popular subtopics (in terms of number of papers). The full results of this example study, however, are available at \url{rexplore.kmi.open.ac.uk/data/edam} and on Zenodo\footnote{\url{https://doi.org/10.5281/zenodo.2653924}}   and can be reused for supporting a more in-depth analysis of the field. 

\begin{table}[htpb]
\centering
\includegraphics[scale=0.82]{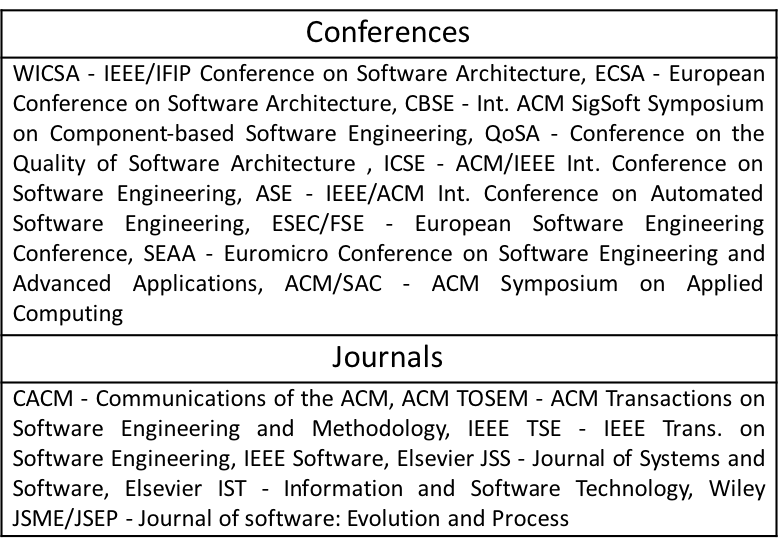}
\caption{List of venues used for the DSA-MV dataset.}
\label{tbl:venuetable}
\end{table}

\begin{figure}[htpb]
\centering
\includegraphics[scale=0.35]{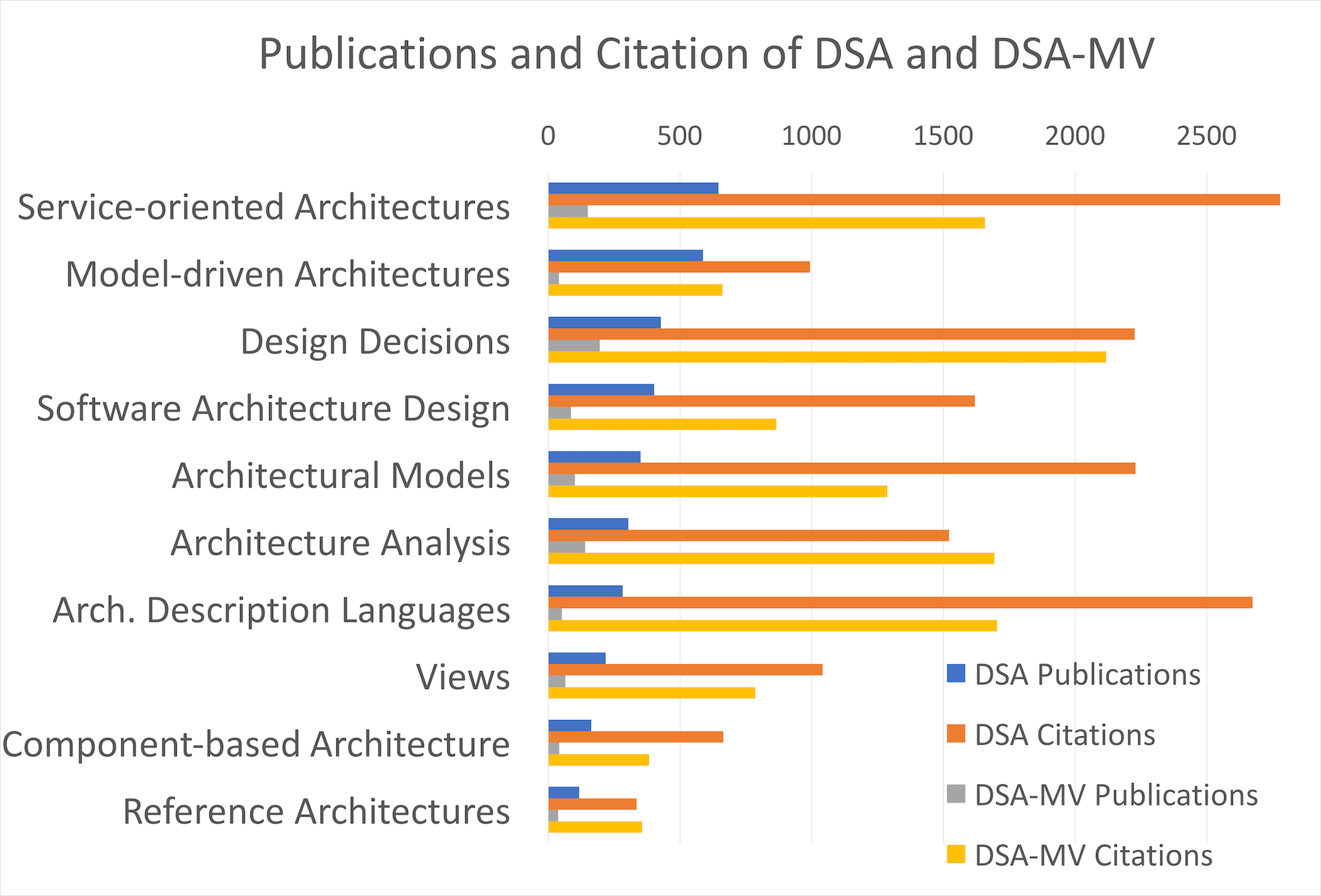}
\caption{Number of publications and citations of the main topics in DSA and DSA-MV.}
\label{fig:study2}
\end{figure}

Figure~\ref{fig:study2} displays the number of publications and citations associated with the most popular sub-areas of SA. The papers in DSA yield on average $4.8 \pm 2.1$ in citations versus the $13.6 \pm 7.0$ citations of those in DSA-MV. Reasonably, this tendency suggests that the papers published in the main SA venues tend to be more recognized by the research community.

\begin{figure}[htpb]
\centering
\includegraphics[scale=0.30]{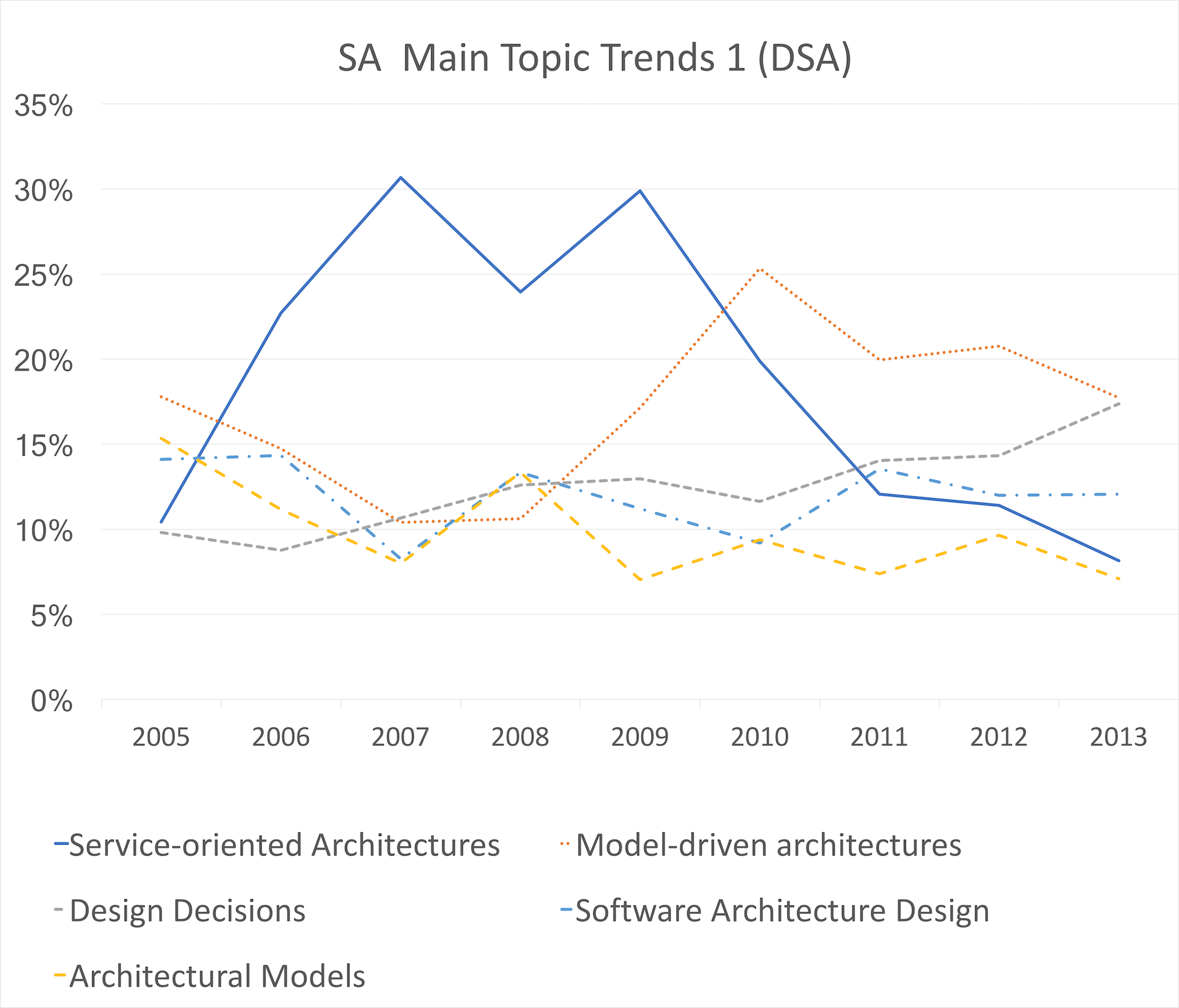}
\includegraphics[scale=0.30]{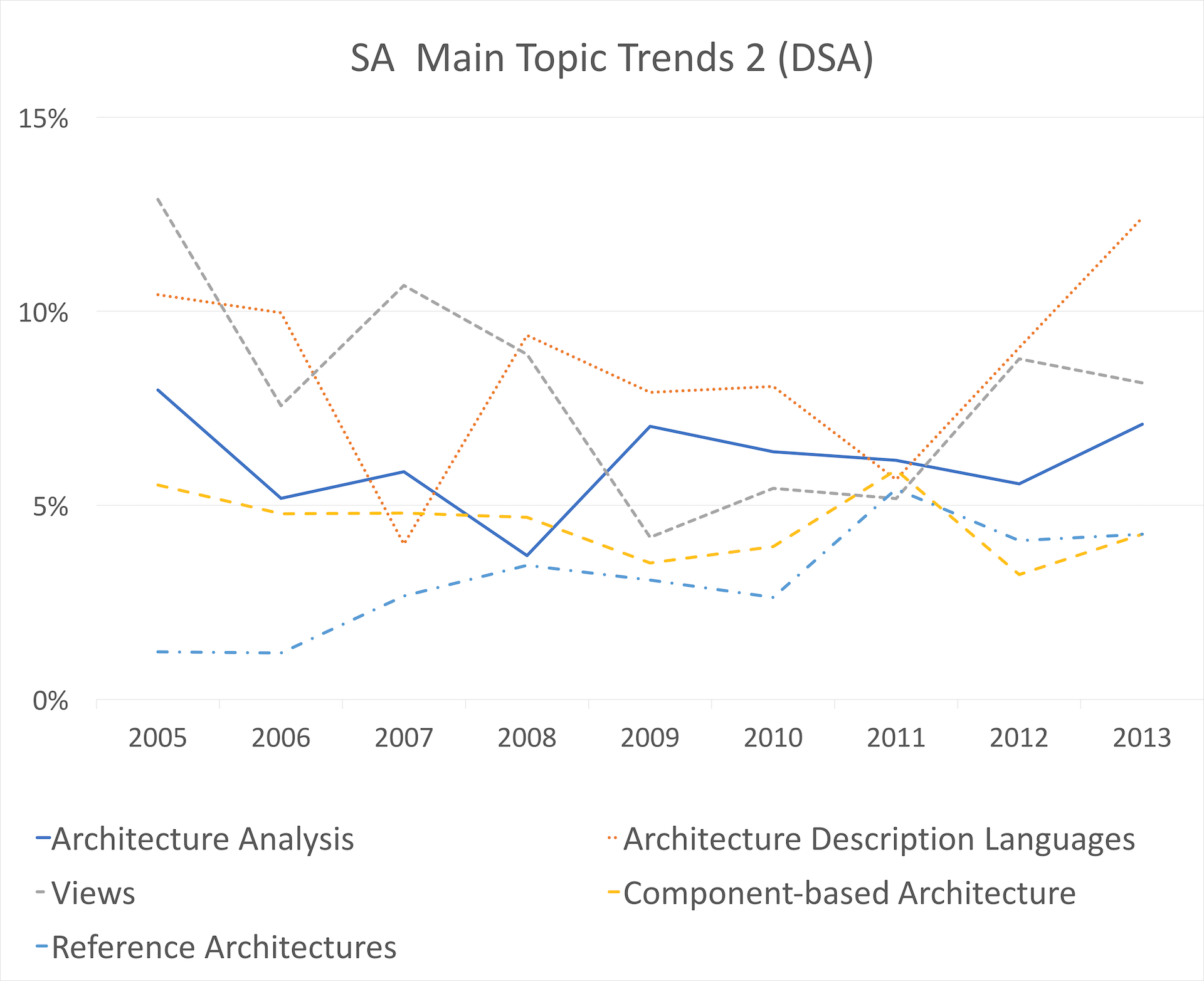}
\caption{Number of publications of the top ten main topics in DSA over time.}
\label{fig:study3}
\end{figure}

Figure~\ref{fig:study3} shows the percentage of papers published over time in the main topics within SA. We focus on the 2005-2013 period, since in this interval the number of publications is high enough to highlight the topic trends.

Software-oriented Architectures appears to have been the most prominent topic before 2009, while from 2010, Model-driven Architectures appears to be the most popular topic in this dataset. We can also appreciate the rising of Design Decisions, that seems the most significant positive trend of the last period together with Architecture Description Languages.

\begin{figure}[htpb]
\centering
\includegraphics[scale=0.13]{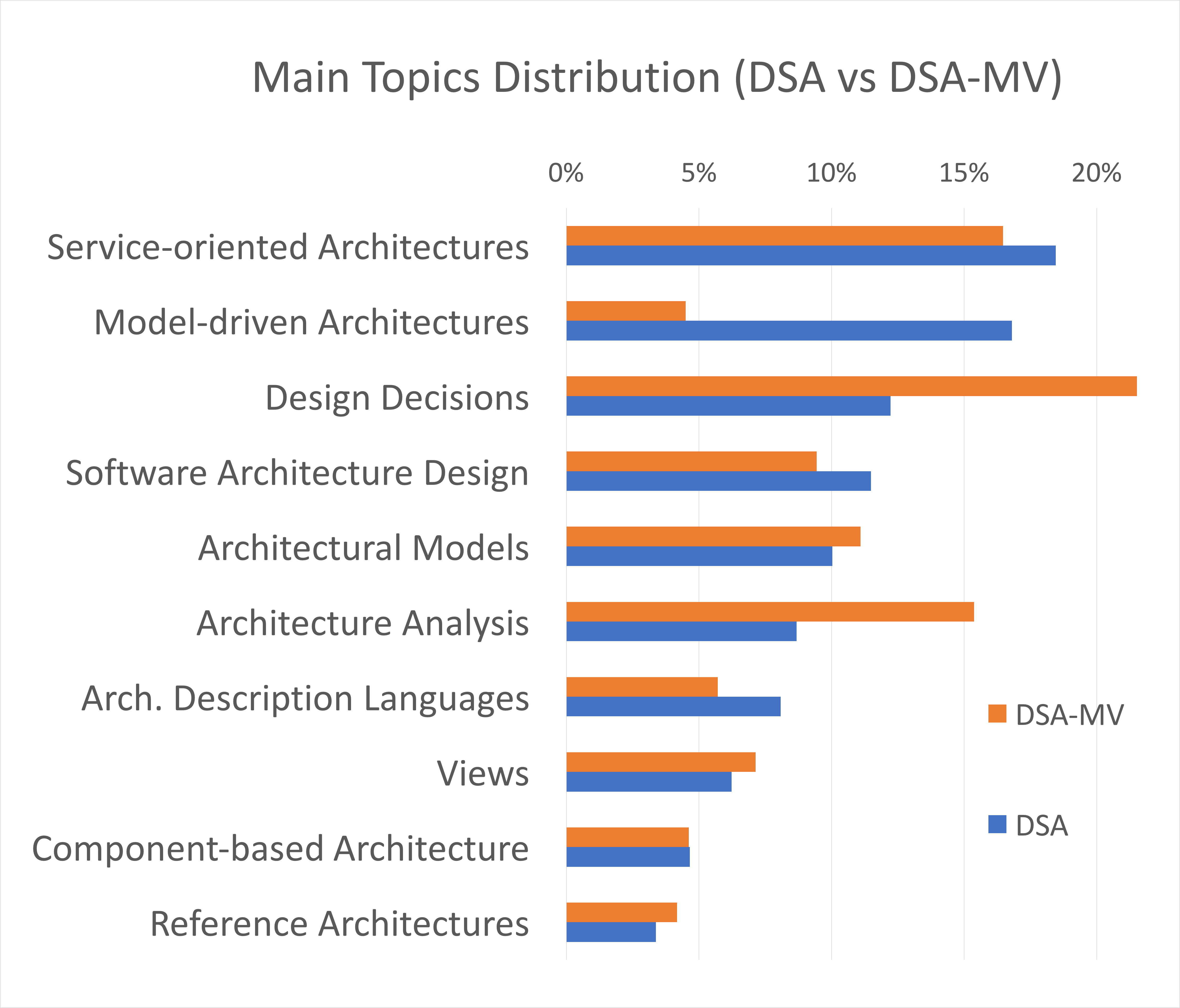}
\caption{Comparison DSA and DSA-MV in terms of topic distribution. The percentage value refers to the ratio between the number of publications in a topic and the total publications in the ten main topics.}
\label{fig:study5}
\end{figure}

\begin{figure}[htpb]
\centering
\includegraphics[scale=0.30]{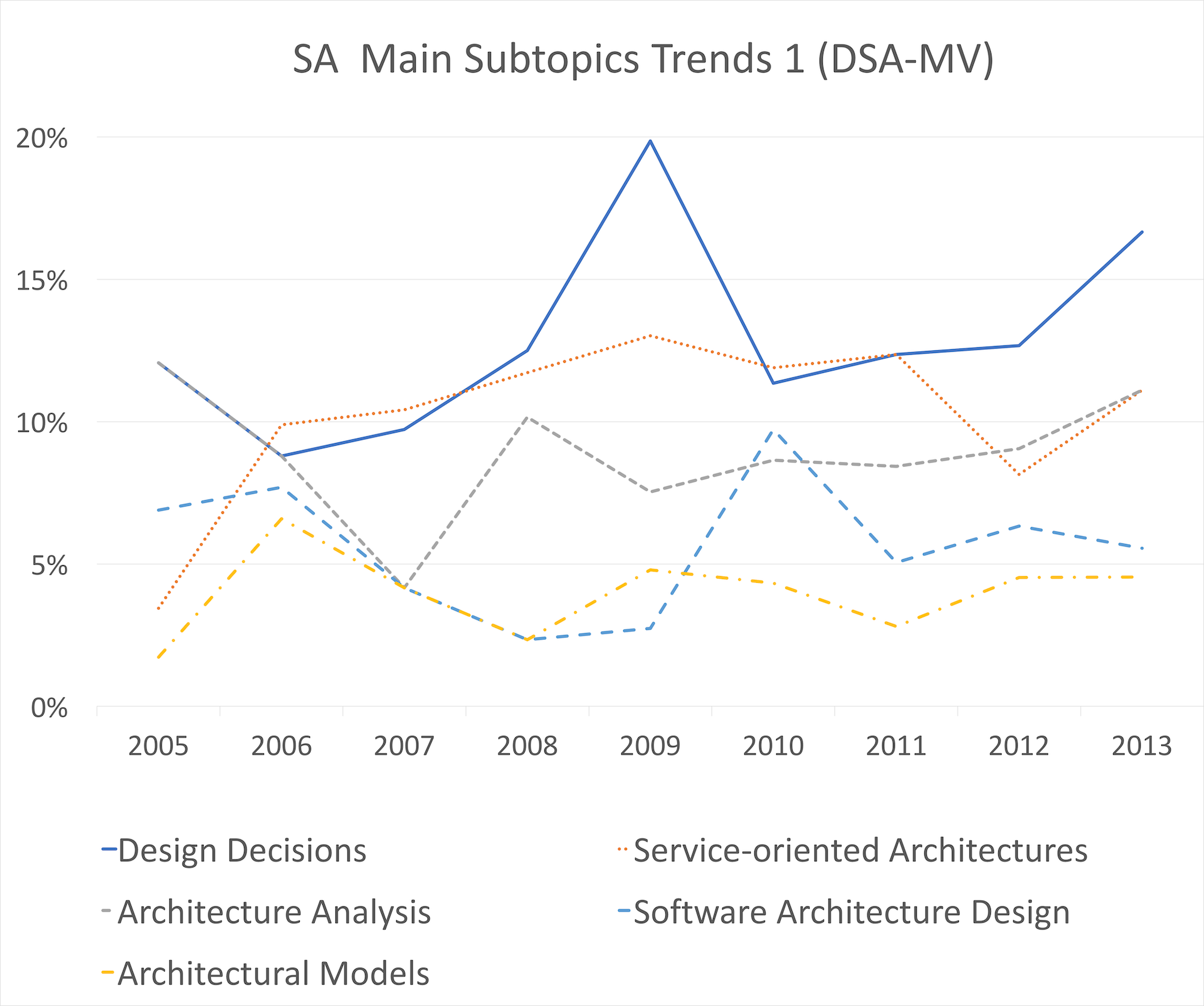}
\includegraphics[scale=0.30]{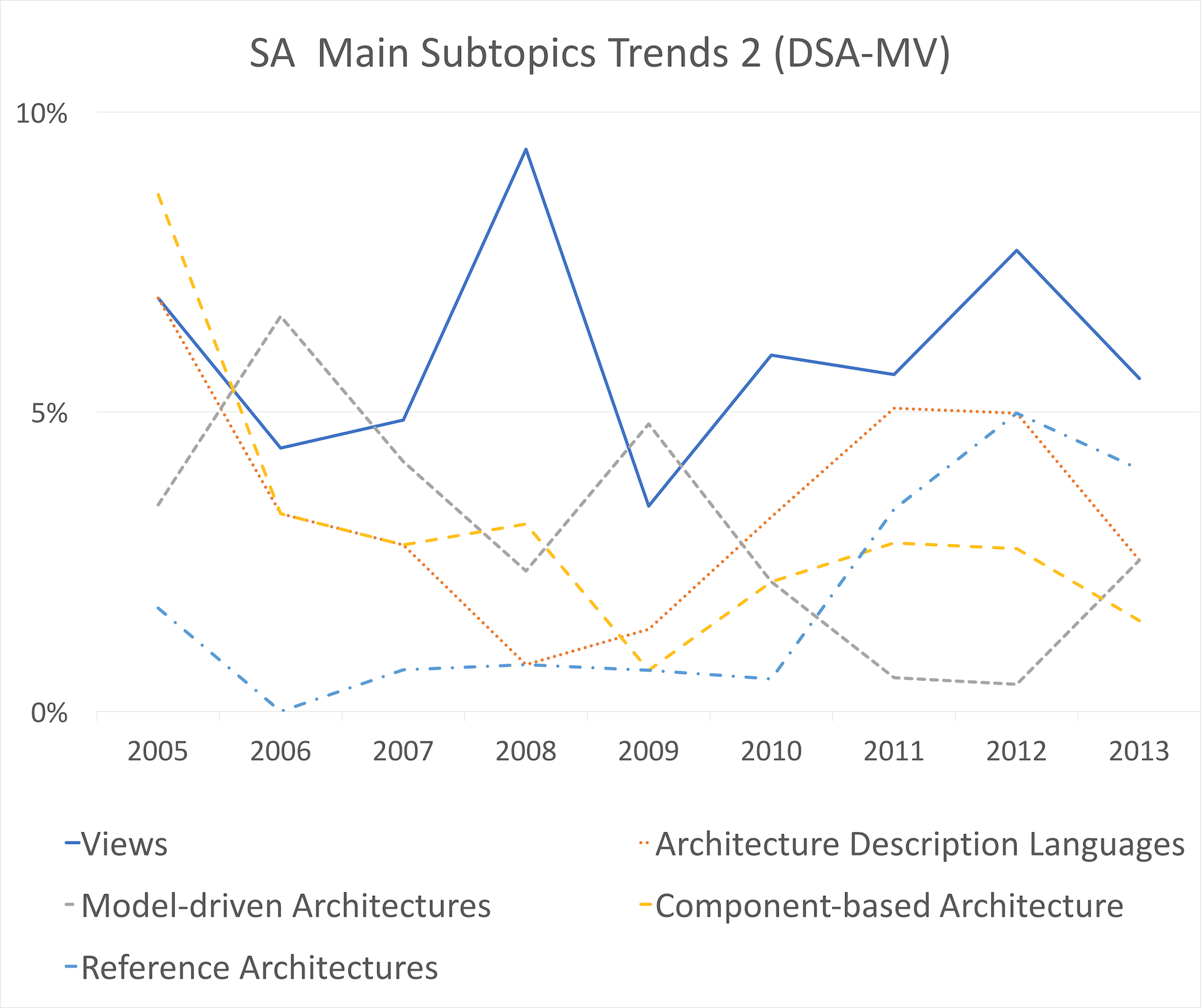}
\caption{Number of publications of the top ten main topics in DSA-MV over time.}
\label{fig:study6}
\end{figure}

Interestingly, the dataset regarding the main venues (DSA-MV) exhibits some different dynamics. Figure~\ref{fig:study5} highlights the difference between DSA and DSA-MV by showing for each topic the ratio between its number of publications and the total publications in the ten main topics. The research areas of Design Decisions and Views appear much more prominent in the main venues, while Model-Driven Architectures and Architecture Analysis are more popular in DSA. We can further analyze these differences by considering the main trends of the DSA-MV dataset, displayed by Figure~\ref{fig:study6}. 
The trend of Design Decisions in DSA-MV mirrors the one exhibited in DSA, both growing steadily from 2010. Conversely, Service-oriented Architectures, which has a negative trend in DSA, remains stable in DSA-MV.

\section{Evaluation and Discussion}  \label{sec:discussion}
In the following, we reflect on this preliminary application of EDAM. This section includes 1) a evaluation of our method versus six human annotators, 2) a comparison of several approaches for classifying primary studies, 3) an analysis of EDAM limitations, 4) a discussion about the implications for systematic mappings in Software Engineering, and 5) a discussion on how to reuse EDAM for other SRs.

\subsection{Evaluation of the primary study classification} \label{sec:eval1}

The most critical step of EDAM is the classification of primary studies. When these are correctly associated to the relevant topics, the subsequent analysis presents a realistic assessment of the landscape of the studied research field. Thus, even if working on a large number of papers can alleviate the weight of some minor misclassification mistakes, we need to be able to trust that the automatic classification process will obtain an accuracy similar to that yielded by human annotators. 

We evaluated the ability of EDAM to correctly discriminate between different topics in the field of Software Architecture by (1) randomly selecting a set of 25 papers in the DSA dataset, (2) classifying them both with  EDAM and with six human experts (researchers in the field of SA), and (3) comparing the results. For simplifying the task and allowing to compare the annotation algorithmically, we first selected five unambiguous categories from the main topics of SA: Design Decisions, Service-oriented Architectures, Model-driven Architectures, Architecture Description Languages, and Views. For each category, we randomly selected from the DSA dataset five primary studies that were classified by EDAM exclusively under that topic, for a total of 25 papers. These papers were described in a spreadsheet by means of their title, author list, abstract, and keywords. The human experts were given this spreadsheet and asked to classify each paper either with one of the five categories or with a ``none of the above'' tag. We then compared the seven annotation sets produced by the six human experts and by EDAM, considered as an additional annotator\footnote{The material and the results of the evaluation are available at \url{https://doi.org/10.5281/zenodo.2653924}}. 

Table~\ref{tbl:agreement} shows the agreement between the annotators. It was computed by calculating the ratio of papers which were tagged with the same category by both annotators. EDAM has the highest average agreement and it also yields the highest agreement with three out of six users. User5 does even better in this regards and has the highest agreement with four  annotators.

The chi-square test run on the human users shows that their behaviours are significantly different ($p=0.017$). However, if we group together users $\{2,3,5,6\}$ and users $\{1,4\}$, we find no significant differences in the behaviour within each group ($p=0.81$, $p=0.38$, good intra-group agreement), while there are between the two groups ($p = 0.0007$). 

Interestingly, users $\{1,4\}$ were two students at the beginning of their PhD, hence still relatively new to the domain. This could suggests the importance of considerable domain experience for this task. EDAM exhibits a behaviour consistent with the most senior group, from which it is not significantly different ($p=0.77$).

\begin{table}[htpb]
\centering
\includegraphics[scale=0.62]{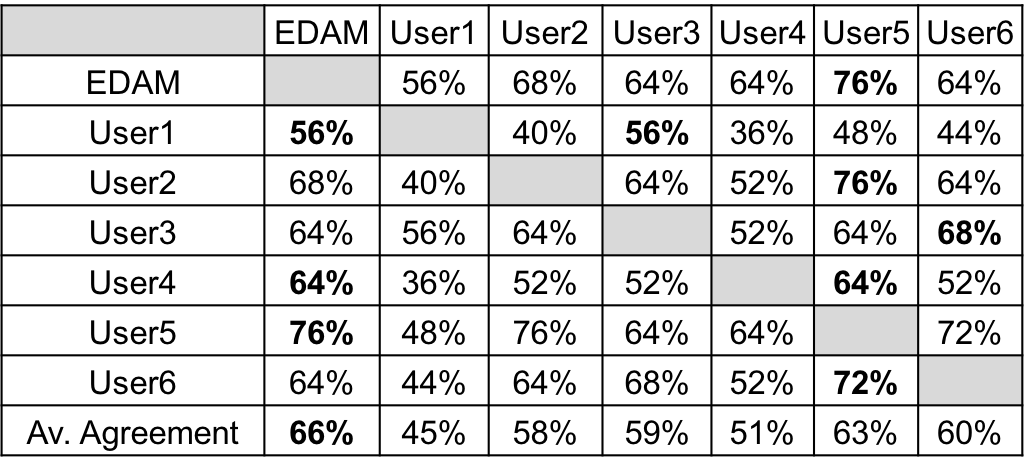}
\caption{Agreement between annotators (including EDAM) and average agreement of each annotator. In bold the best agreements for each annotator.}
\label{tbl:agreement}
\end{table}	

\begin{figure}[htpb]
\centering
\includegraphics[scale=0.120]{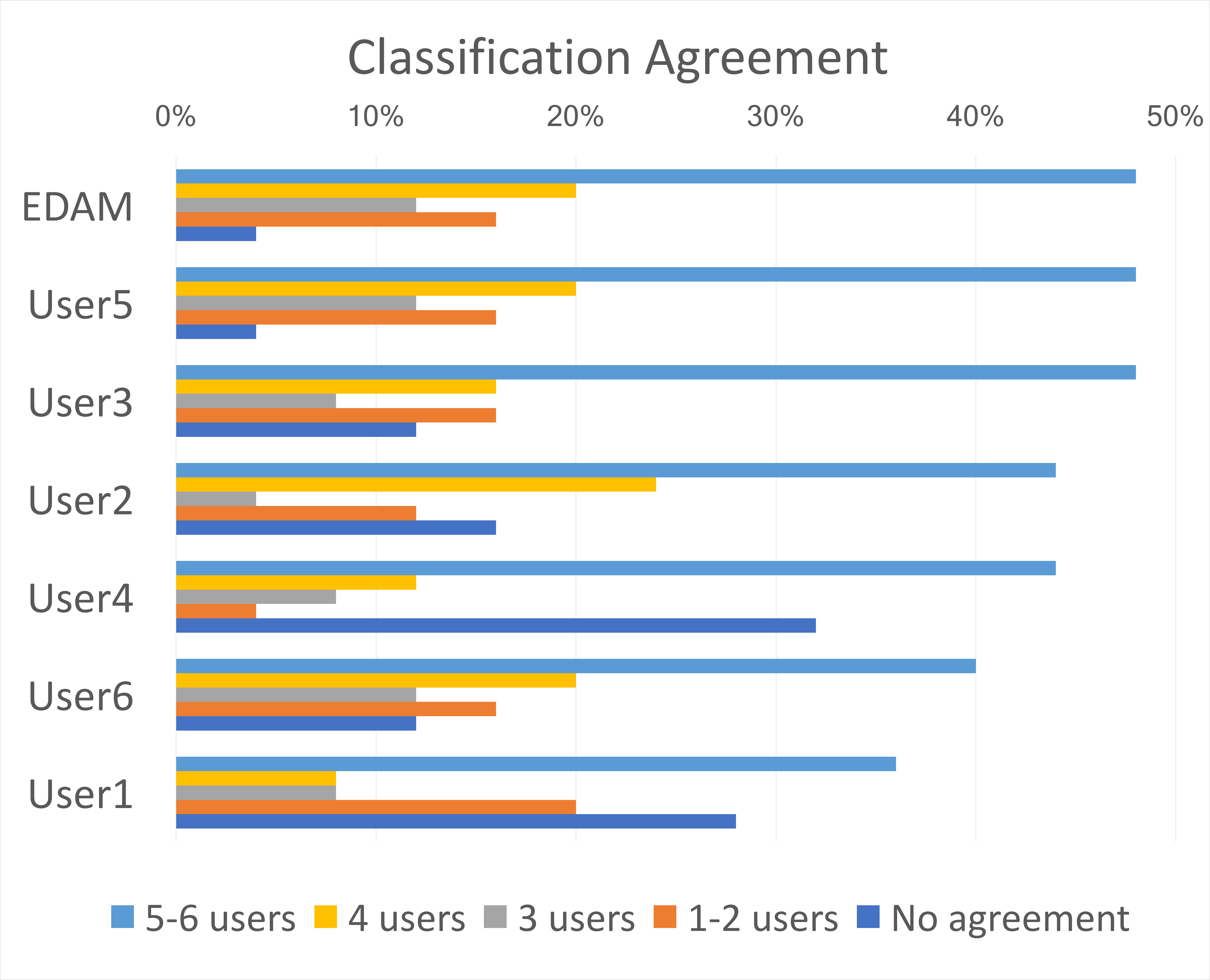}
\caption{Percentage of annotations that agree with other $n$ annotators.}
\label{fig:evalchart}
\end{figure}	

As anticipated, a good way to measure the performance of annotators is their agreement with the majority of other expert users. Figure~\ref{fig:evalchart} shows the percentage of annotations of each annotator that agree with other $n$ annotators. EDAM agrees with four out of six human annotators for 68\% of the studies, it agrees with at least three of them for 80\% of the studies, and it agrees with at least one of them for all the studies but one. Indeed, the categories generated by EDAM coincide with the ones suggested by the relative majority of users in 84\% of the cases. Therefore, EDAM's performance is comparable to the performance of the two annotators (User5 and User3) with the highest agreement with the user majority. 

We further confirmed these findings by computing the Cohen's kappa between each couple of annotators and between each of the annotators and EDAM. The inter-annotator agreement was $0.57$, typically indicating a moderate agreement \citep{landis1977measurement}. The average agreement of EDAM with the annotators was $0.58$, confirming that this method performs in line with the annotators. In addition, we note that EDAM always agrees with the majority for the studies in which no more than one annotator disagrees. It thus seems to  perform well in handling simple not-ambiguous papers, that nonetheless human experts may sometimes get wrong.

In conclusion, this study suggests that the EDAM classification step generates annotations that agree with the majority of human experts and are not statistically different from the ones produced by the senior group.

Naturally, EDAM performance may change according to the quality of the ontology and the domain knowledge of the human users that refined it. EDAM is not an alternative to human experts, rather a methodology that allows humans to annotate on a larger scale, by defining a sound domain knowledge and a mapping function. However, this preliminary example application already shows very promising results.

\subsection{Comparison of Classifiers for Primary Studies}\label{sec:eval2}

EDAM can adopt many different approaches for automaticaly classifying primary study. The choice of the method ultimately depends on its affectiveness of the avaliable approaches on the domain under analysis and on the preferred precision-recall tradeoff. For instance, in a prevalently automatic mapping study on a large set of papers, precision is of paramount importance, and the missed topics may be compensated by the numerosity of the sample. Conversely, when the results are validated in some way by human experts (e.g.,~\citep{osborne2016automatic}) or when the goal is to detect emerging trends that may appear in few publications, a better strategy may be to sacrifice some precision for producing a more comprehensive set of topics.

In this section we compare several approaches that can be used with EDAM to automatically classify studies according to a taxonomy of research topics, and discuss their tradeoff. We focus on unsupervised approaches, since typically the authors of a systematic review do not have the resources to prepare a large gold standard to train a supervised classifier on a potentially new taxonomy. 

We evaluated seven alternative approaches on a gold standard of 70 papers~\citep{salatino2018csoc2} within the fields of Semantic Web (23 papers), Natural Language Processing (23), and Data Mining (24). These papers were selected by retrieving the most cited papers from Microsoft Academic Graph containing in the title or the abstract those relevant fields. Each paper was annotated by three domain experts (for a total of 21 different annotators) and was associated with $14.4\pm 7.0 $ topics using majority vote in case of disagreement. The inter-annotator agreement was  $0.45\pm 0.18$ according to Fleiss' Kappa, which indicates a moderate inter-rater agreement~\citep{landis1977measurement}. The data produced in the evaluation and the Python implementation of the approaches are available at \url{https://cso.kmi.open.ac.uk/cso-classifier/}. A more comprehensive version of this analysis, with additional baselines that are out of scope for this paper (e.g., not producing a set of pre-defined categories), is available in ~\citet{salatino2018csoc2}.

All the tested classifiers analysed the title and abstract of the 70 papers and assigned them with a set of topics drawn from the Computer Science Ontology (CSO), a recently released taxonomy of research areas~\citep{salatino2018computer}. This knowledge base covers well the three mentioned fields, including a total of 35 sub-topics for the Semantic Web, 173 for Natural Language Processing, and 396 for Data Mining. LDA100, LDA500, and LDA1000 are based on a Latent Dirichlet Allocation (LDA) model ~\citep{blei2003latent} trained over 4.6M papers in Computer Science from Microsoft Academic Search. Specifically, LDA100 used a model trained with 100 topics, LDA500 on 500 topics, and LDA100 on 1,000 topics. Similarly to~\citep{bhatia2016automatic}, these classifiers generate a set of CSO topics from the LDA topics by first producing a set of topics with a probability of at least $j$ and all their terms with a probability of at least $k$. Then they map these terms to CSO by returning all CSO topics having Levenshtein similarity higher than 0.8 with them. We performed a grid search for finding the best values of $j$ and $k$ on the gold standard and report here the best results of each classifier in term of F-measure.

TF-IDF produces a ranked list of terms using TF-IDF~\citep{ramos2003using} (the IDF was computed on the same set adopted for LDA) and all the CSO topics having Levenshtein similarity higher than 0.8 with the first 30 terms. 

 Direct Mapping (DM) is the approach used for the implementation of EDAM described in step 6 of  Section~\ref{sec:methodologyimplementation}. This same method was also used by the first version of the Smart Topic Miner (STM) \citep{osborne2016automatic}, the system adopted by Springer Nature to classifying proceedings in the field of Computer Science. 
 It returns all topics that explicitly appear in the papers or that are entailed by the ones appearing in the paper according to the ontology.
 
 The CSO Classifier v.1 (CSO-C1) is an unsupervised approach presented in~\citet{salatino2018csoc} that extracts a combination of n-grams (unigrams, bigrams, and trigrams) from the text and returns all the topics that have a Levenstein similarity higher than $t$ ($t=0.94$ as in the implementation reported in~\citet{salatino2018csoc}). Finally, the CSO Classifier v.2 (CSO-C2)~\citep{salatino2018csoc2} is a recent evolution of the previous classifier, which  uses part-of-speech tagging to identify promising terms and then exploits word embeddings to infer semantically related topics that may not explicitly appear in the paper. This solution has also been adopted by the current version of the Smart Topic Miner \citep{salatino2019improving}. 
 
 \begin{table}[htpb]
\centering
\includegraphics[scale=1]{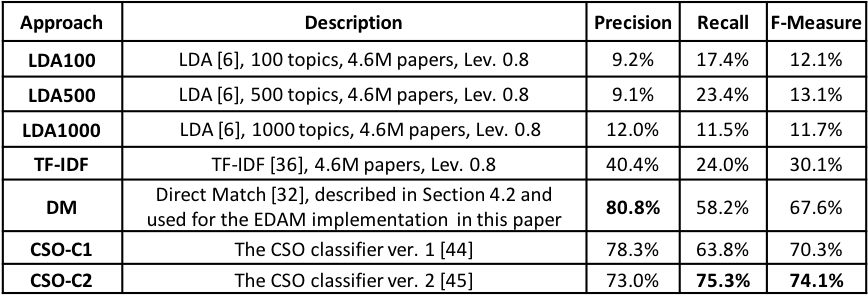}
\caption{Values of precision, recall, and F-measure for the seven classifiers. In bold the best results.}
\label{tbl:eval2}
\end{table}	

 Table \ref{tbl:eval2} reports on the resulting values of precision, recall and F-measure. The approaches based on LDA performed quite poorly. An analysis of the results revealed that the topics returned by the models are both noisy and coarse-grained, often clustering together distinct topics from CSO. Indeed, while LDA works quite well at identifying the main topics of a large collection of documents, it does not traditionally perform equally well when characterizing  specific research topics, which may be associated with a relatively low number of publications (50-200), as discussed in~\citep{osborne2012mining}. The approaches based on TF-IDF worked slightly better, yielding an F-measure of 30.1\%.
 DM, used in our exemplary EDAM implementation, yielded the best precision of all the approaches (80.8\%). Indeed, this method focuses on topics that are explicitly mentioned in the text, which tends to be very relevant. However, this solution naturally obtains a relatively low recall of 58.2\%. CSO-C1, which expands the set of terms by considering string similarity, obtained a better recall (63.8\%) but a lower precision (78.3\%). Finally, CSO-C2 yielded the best performance in term of both recall (75.3\%) and F-measure (74.1\%).
 
 DM performed significantly better ($p<10^{-7}$ with the McNemar's test) than the first four approaches. In turn, CSO-C2 performed  significantly better($p<10^{-7}$) than DM and CSO-C1.
 
\subsection{Limitations}\label{sec:limitations}
In this section we discuss EDAM limitations based on the categorization given in \citet{wohlin2012experimentation}.

For {\em internal validity} we have identified two main threats that regard the generation of a reliable ontology, which is key to select relevant studies that directly fulfill the selection criteria (and hence correspond to the primary studies for the study at hand). In particular:
\begin{description}
\item{\bf Ontology learning (step 3): hierarchy is important.} The domain ontology, automatically inferred by the ontology learning technique, is structured hierarchically. Therefore, an area marked as {\em subarea} (e.g., architecture description languages) is subsumed by the previous area at the upper level of the taxonomy (e.g., Software Architecture). {\em Deeper hierarchies bring finer-grained topics, and therefore a higher precision in the classification process}. 

During the application of ontology learning techniques to various research areas (not reported in this paper for the sake of brevity) we found that current ontology learning methods usually identify only mature (in terms of number of publications) research areas. Emerging topics may be excluded, thus reducing the granularity of recent fields' ontologies.

To alleviate this problem, human experts may be asked to manually identify the most recent areas and to possibly adopt ontology forecasting techniques \citep{cano2016ontology}. Therefore, the role of experts in improving the quality and deepness of the hierarchy is indeed critical. For the sake of this study, aimed at showing the advantages of automation, the relatively small number of experts was acceptable. However, a larger and more diversified pool of experts should be involved when the research area under investigation is broader.
	
\item{\bf Ontology refinement (step 4): experience matters.} As illustrated in Figure \ref{fig:edam}, EDAM requires human expertise to refine the automatically generated ontology (step 4). This task is not always straightforward, since humans can have different views on the foundational conceptual elements characterizing a certain discipline. Those differences may be related to many factors, such as the researcher's exposure to the research area under investigation, seniority, broad vs. specialized knowledge on specific sub-disciplines. Our preliminary experiments allow us to conclude that senior domain experts, with a mature yet wide view on the research area under investigation, should be selected to minimize this threat. 

\end{description}

The main threats for {\em external validity} regard the practical exploitation of EDAM. In particular:
\begin{description}
\item{\bf Scholarly dataset: different research areas require different datasets.} This paper reports on our experience with EDAM's application to the Software Architecture research area.
Since the domain of Software Engineering is well represented in the Scopus dataset, we are not facing generalizability issues. However, moving to a totally different domain would require taking into account (assuming to have access to) different scholarly datasets. 

Unfortunately, finding up-to-date datasets of scholarly data covering the field under analysis is not always easy and this could be a threat to our approach. Nonetheless, the movement toward open access is helping in mitigating this issue by making available a variety of datasets containing machine-readable data about scientific publications, e.g.,
Microsoft Academic Graph\footnote{\url{https://academic.microsoft.com/}}~\citep{sinha2015overview},
CORE\footnote{\url{https://core.ac.uk}}~\citep{knoth2012core},
OpenCitations\footnote{\url{http://opencitations.net/}}~\citep{peroni2015setting},
DBLP\footnote{\url{http://dblp.uni-trier.de/}}~\citep{ley2009dblp},
Bio2RDF\footnote{\url{bio2rdf.org/}}~\citep{belleau2008bio2rdf},
ScholarlyData.org\footnote{\url{http://www.scholarlydata.org/}}~\citep{nuzzolese2016conference},
Nanopub.org\footnote{\url{http://nanopub.org/}}~\citep{kuhn2013broadening},
Semantic Scholar\footnote{\url{https://www.semanticscholar.org/}}, 
and others.

\item{\bf Tool support: closed-source tools.} EDAM is making use of some closed-source, proprietary tools for running some of the tasks. This may reduce the application of our approach from other research groups. In order to mitigate this threat, we are planning to release a web service accessible by other colleagues interested to carry out an EDAM study. 

\item{\bf Research Questions: some may not be automatized.} Many research questions that are typical of mapping studies can be answered by producing relevant analytics~\citep{wohlin2012experimentation}, e.g., by counting the number of publications, authors, and venues associated with certain topics in subsequent years. However, some more complex research questions may still require domain experts to manually analyse the relevant studies, e.g., for classifying them in categories that a state of the art classifier would be unable to detect with good accuracy. This is an inherent limitation of the methodology. Nonetheless in many of these cases a preliminary classification by an automatic system may still alleviate the expert work load, e.g., by reducing the set of publications that need to be manually analysed. In addition, the performance of entity extraction and linking tools is steadily improving~\citep{augenstein2012lodifier, gangemi2017semantic,rizzo2012nerd}, allowing to extract increasingly better representations of research knowledge from scientific articles. Therefore, the number of research questions that can be addressed algorithmically may increase over the following years. 

\end{description}

\subsection{Implications for Systematic Mappings} \label{sec:implications}

There are a few implications that can potentially change the way we perform systematic mapping studies in Software Engineering. As mentioned in Section~\ref{sec:method}, these implications regard:
\begin{description}

\item{\bf Scalability: size does not matter anymore.} EDAM can process a potentially endless set of publications. This allows e.g., mapping studies to be based on {\em all} relevant primary studies, previously scoped down due to the fact that humans could not manually process hundreds or thousands of papers. 

\item{\bf Objectivity: the automatic classification is less biased.} The automatic classification of primary studies does not suffer from the biases of specific human annotators. Nonetheless, the quality of the classification appears on par with the one produced by the human annotators. 

\item{\bf Reproducibility: study duplication and extension is easy.} Thanks to EDAM, replicating or extending studies, either by the same researcher or by someone else, requires simple tuning, e.g., to extend the publication period, or to select different views illustrating the publication trends of interest.

\item{\bf Granularity of the study: zooming-in and -out is simpler.} Thanks to the fact that the selection and classification of primary studies is based on an domain ontology, and of course to automation, EDAM allows to tune the depth of the classification the researcher desires in a given research area. Such tuning just requires setting the level of categories and sub-categories to be included in the classification, and then re-run the methodology.
\end{description}

\subsection{Reusing EDAM for other Systematic Reviews} \label{sec:selection}

\begin{figure}[ht]
\centering
\includegraphics[width=\textwidth,height=.95\textheight,keepaspectratio]{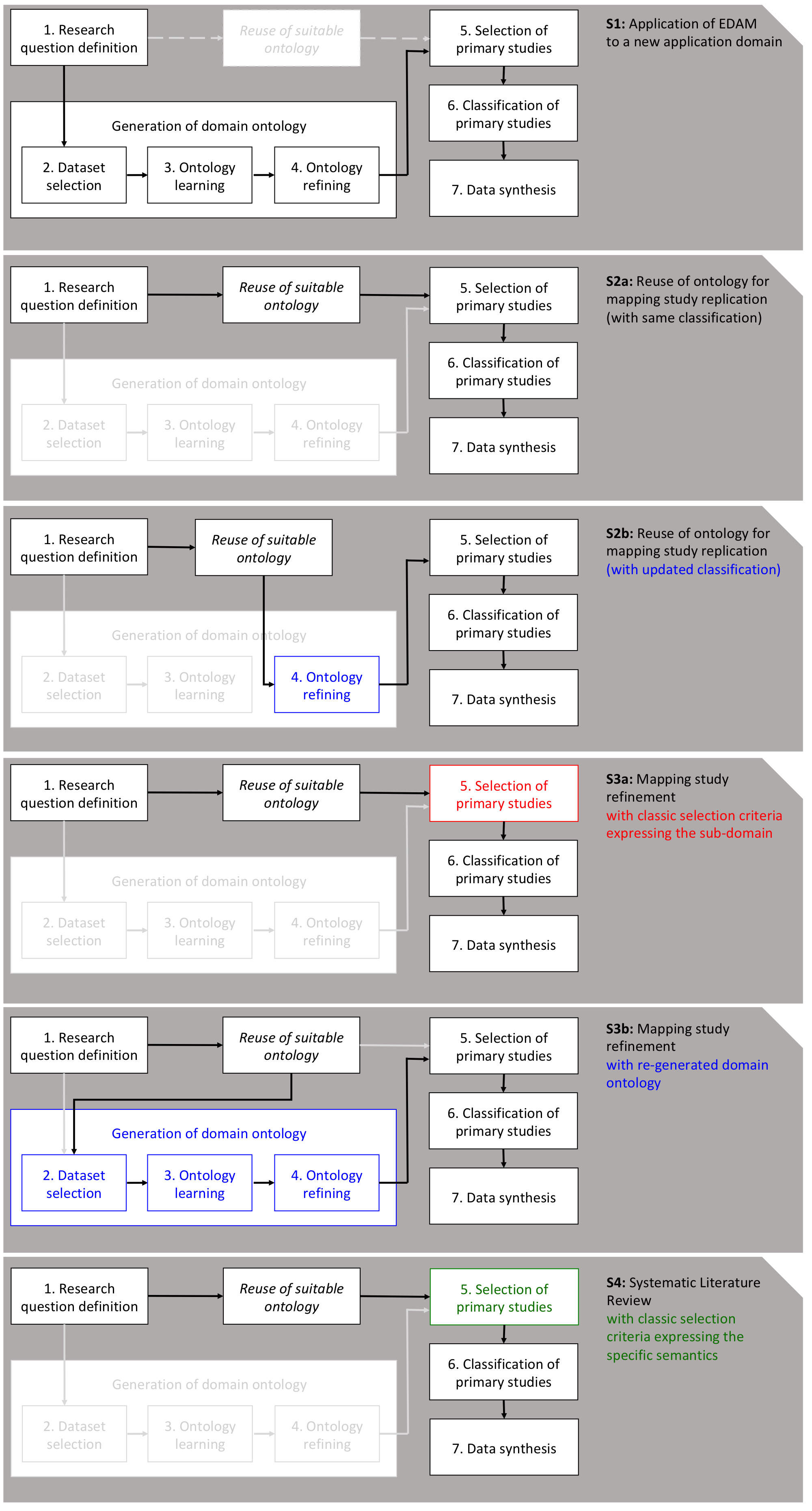}
\caption{Possible EDAM applications.}
\label{fig:scenarios}
\end{figure}

EDAM can be applied to any domain of interest and for different types of studies. The scenarios that we envisage are discussed below and illustrated in Figure~\ref{fig:scenarios}. They are: S1) Application of EDAM to a {\em new} application domain, S2) Mapping study {\em replication}, S3) Mapping study {\em refinement}, and S4) {\em Systematic literature review}.

\begin{description}
\item[{\bf Application of EDAM to a new application domain (S1).}] In the basic scenario ({\bf S1}), the ontology for the new application domain is not yet available. In this case, the complete process illustrated in Figure~\ref{fig:edam} (and emphasized in Figure~\ref{fig:scenarios}.(S1)) shall be applied. This is the scenario followed in the work presented in this article. It is applicable while investigating a new domain notwithstanding its specific characteristics.
\end{description}

If instead a researcher wants to perform a SR in a domain for which the ontology already exists (scenario S2), such generated domain ontology can be {\em reused} in the following two ways, depending on the specific study goal:

\begin{description}

\item[{\bf Mapping Study Replication (same classification, S2a).}] Suppose we wa\-nt to replicate a pre-existing EDAM mapping study conducted at time t$_{0}$, in order to update the list of primary studies and related analysis at time t$_{1}$ (e.g., update in year 2020 the study on Software Architecture presented in this paper). In this case, we can directly reuse the previously generated ontology (cf. Figure~\ref{fig:scenarios}.(S2a)). The list of (updated) primary studies can be automatically re-calculated (in step 5) and used (in step 6) for classification and analysis purposes. Notice, however, that this scenario does not address the potential need to {\em update} the list of topics. Such a scenario is covered below.

\item[{\bf Mapping Study Replication (updated classification, S2b).}] Differently from 
scenario S2a, we may be interested to replicate a pre-existing study {\em and} also include any new topics that may have emerged in the period between time t$_{0}$ and time t$_{1}$ (e.g., updating this study in year 2020 while including new topics appeared after this study). This need requires an update of the domain ontology; therefore, the process in Figure~\ref{fig:scenarios}.(S2b) must be run from step 4 onward.

\end{description}

Another scenario (S3) accommodates the case in which we want to {\em refine} the classification and analysis conducted as a mapping study. In the current approach, as shown in the Software Architecture domain scenario, step 5 in Figure~\ref{fig:edam} returns a set of primary studies that can be further classified into sub-domains (e.g., Architectural Styles, being one element of our ontology, can be further refined to discover all the papers that cover selected styles). We identify two sub-scenarios in order to provide a refinement of sub-domains contents:

\begin{description}

\item[{\bf Mapping Study Refinement with classic selection criteria (S3a).}] 
In t\-h\-is scenario, one may classify the articles into sub-domains of interest by applying the inclusion and exclusion criteria \citep{kitchenham2007guidelines} to the primary studies selected in step 5 of EDAM. For example, knowing that Publish-Subscribe, Client-Server, and Event-driven are sub-domains of Architectural Styles, we introduce selection criteria to position Architectural Styles articles into those categories. This approach allows us to zoom into a specific sub-domain of interest and extract the articles fitting in the specific target sub-domain. 
\item[{\bf Mapping Study Refinement with re-generated domain onto\-lo\-gy (S3b).}] The selected sub-domain of interest may contain hundreds of papers (for example, the Design Decisions sub-domain in our study includes 428 papers). Consequently, applying the selection criteria reported in scenario S3a may be cumbersome, requiring the manual analysis of most of those papers. Alternatively, the researcher may execute an additional round of steps 2-4 to refine the domain ontology for the specific sub-domain (cf. Figure~\ref{fig:scenarios}.(S3b)). This scenario is similar to S1, but applied to a specific sub-domain of interest.

\end{description}

A fourth scenario sees the researcher is interested to run a systematic literature review (SLR) on specific research questions:

\begin{description}

\item[{\bf Systematic Literature Reviews (S4).}] In step 5 (cf. Figure~\ref{fig:scenarios}.(S4)), given the list of primary studies generated based on the existing ontology, we may run the {\em classic} SLR approach \citep{kitchenham2013systematic} to select those papers that fit with the research questions of interest. Differently from scenario S3a, S4 adds the semantics beyond the definition of the domain, and encapsulated into the research questions and the corresponding selection criteria. E.g., given the list of all studies on Software Architecture styles, one may want to perform an SLR to analyze those approaches that are adopted in industrial settings.

\end{description}

\section{Conclusions and Future Work} \label{sec:conclusion}
In this paper we have presented EDAM, an expert-driven automated methodology to assist systematic reviews. Its application to the Software Architecture research area shows preliminary and very promising results.

Motivated by the large amount of time and effort needed by classic methodologies to select and classify the primary studies, EDAM offers benefits that can help SE researchers to dedicate most of their time to the most cognitive-intensive tasks like e.g., interpretation of the trends and extraction of lessons and research gaps.

Additional benefits have been emphasized in Section~\ref{sec:generalmethodology} (after presenting EDAM) and Section~\ref{sec:implications} (discussing implications for systematic mappings). Among the benefits we mention the great potential for re-using EDAM and in particular domain ontologies and functions to build a shared framework helping the research community at large. Much can be done in this direction.

Our next step is to complement EDAM with automated forward snowballing to further reduce the effort for identifying relevant primary studies. 
With the same goal, we are planning to investigate other possible data synthesis techniques through machine learning techniques or the (manual) intervention of human experts. Last, but most important for us, we plan to reconstruct the 25 years of the Software Architecture body of knowledge by fully exploiting EDAM automation and human expertise.

\section*{Acknowledgments} \label{sec:acknowledgments}
The authors would like to thank  the colleagues which donated their time and expertise by contributing to this study as domain experts and/or annotators: Paris Avgeriou, Barbora Buhnova, Rafael Capilla, Jan Carlson, Ivica Crnkovic, John Grundy, Rich Hilliard, Heiko Koziolek, Anton Jansen, Ivano Malavolta, Leonardo Mariani, Marina Mongiello, Matthias Naab, Patrizio Pelliccione, Mohammad Sharaf, Damian Andrew Tamburri, Antony Tang, Jan Martijn van der Werf, Smrithi Rekha Venkatasubramanian, Rainer Weinreich, Danny Weyns, Eoin Woods, and Uwe Zdun.

We also thank  Davide Falessi for reviewing an earlier version of this manu\-scri\-pt, and Elsevier BV for providing us with access to its large repository of scholarly data.



\bibliographystyle{plainnat}
\bibliography{bibliography}        

%

\end{document}